%% file: main.tex
\documentclass{interact}

\usepackage[natbibapa,nodoi]{apacite}
\usepackage[utf8]{inputenc}
\usepackage[automake]{glossaries}
\usepackage{graphicx}
\usepackage{listings}
\usepackage{subfig}
\usepackage{tikz}
\usepackage{xcolor}

\graphicspath{{figures/}}
\let\cite\citep 
\glsdisablehyper
\makeglossaries
\loadglsentries{acronyms}

\begin{document}
\title{Advanced simulation-based predictive modeling for solar irradiance sensor farms}
\author{
\name{José~L.~Risco-Martín\textsuperscript{a}, Ignacio-Iker~Prado-Rujas\textsuperscript{b}, Javier~Campoy\textsuperscript{b}, María~S.~Pérez\textsuperscript{b}, and~Katzalin~Olcoz\textsuperscript{a}\thanks{CONTACT José L. Risco-Martín, Email: jlrisco@ucm.es}}
\affil{\textsuperscript{a}Universidad Complutense de Madrid, Madrid, Spain}
\affil{\textsuperscript{b}Universidad Politécnica de Madrid, Madrid, Spain}
}

\maketitle

\begin{abstract}
  \input{0-abstract}
\end{abstract}


\glsresetall

\section{Introduction and related work}\label{sec:introduction}
\input{1-introduction}

\section{Background}\label{sec:background}
\input{2-background}

\section{System architecture and design}\label{sec:architecture}
\input{3-architecture}

\section{Inference, training, outliers detection and analysis systems}\label{sec:services}
\input{4-services}

\section{Use case}\label{sec:experiments}
\input{5-experiments}

\section{Conclusion and future work}\label{sec:conclusion}
\input{6-conclusion}

\section*{Acknowledgments}
This project has been partially supported by the Education and Research Council of the Community of Madrid (Spain), under research grant S2018/TCS-4423.

\bibliographystyle{apacite}
\bibliography{biblio}

\appendix
\section{DL-based forecasting model}\label{sec:appendix}
\input{7-appendix}

\pagebreak
\printglossary[title=Abbreviation List, type=main, nonumberlist]

\end{document}

%% file: acronyms.tex
\newacronym{ai}{AI}{Artificial Intelligence}
\newacronym{api}{API}{Application Programming Interface}
\newacronym{ann}{ANN}{Artificial Neural Network}
\newacronym{caide}{CAIDE}{Cloud-based Analysis and Integration for Data Efficiency}
\newacronym{cnn}{CNN}{Convolutional Neural Network}
\newacronym{cps}{CPS}{Cyber-Physical Systems}
\newacronym{csi}{CSI}{Clear-Sky Index}
\newacronym{devs}{DEVS}{Discrete Event System Specification}
\newacronym{dl}{DL}{Deep Learning}
\newacronym{dnn}{DNN}{Deep Neural Network}
\newacronym{dt}{DT}{Digital Twin}
\newacronym{eu}{EU}{European Union}
\newacronym{ews}{EWS}{Early-Warning System}
\newacronym{gcp}{GCP}{Google Cloud Platform}
\newacronym{gcs}{GCS}{Ground Control Station}
\newacronym{ghi}{GHI}{Global Horizontal Irradiance}
\newacronym{gpu}{GPU}{Graphics Processing Unit}
\newacronym{gui}{GUI}{Graphical User Interface}
\newacronym{hab}{HAB}{Harmful Algal and Cyanobacterial Bloom}
\newacronym{hil}{HIL}{Hardware In the Loop}
\newacronym{iot}{IoT}{Internet of Things}
\newacronym{ip}{IP}{Internet Protocol}
\newacronym{lstm}{LSTM}{Long Short-Term Memory}
\newacronym{convLstm}{Conv-LSTM}{Convolutional LSTM}
\newacronym{mae}{MAE}{Mean Absolute Error}
\newacronym{mape}{MAPE}{Mean Absolute Percentage Error}
\newacronym{mbse}{MBSE}{Model Based Systems Engineering}
\newacronym{ml}{ML}{Machine Learning}
\newacronym{midc}{MIDC}{Measurement and Instrumentation Data Center}
\newacronym{ms}{M\&S}{Modeling and Simulation}
\newacronym{mse}{MSE}{Mean Squared Error}
\newacronym{mlp}{MLP}{Multilayer Perceptron}
\newacronym[\glsshortpluralkey=NFFs,\glslongpluralkey={Non-Functional Features}]{nff}{NFF}{Non-Functional Feature}
\newacronym{nn}{NN}{Neural Network}
\newacronym{nrel}{NREL}{National Renewable Energy Laboratory}
\newacronym{nwp}{NWP}{Numerical Weather Prediction}
\newacronym{pv}{PV}{Photovoltaic}
\newacronym{pvgis}{PVGIS}{PhotoVoltaic Geographical Information System}
\newacronym{rmse}{RMSE}{Root-Mean-Squared Error}
\newacronym{rnns}{RNNs}{Recurrent Neural Networks}
\newacronym{sdgs}{SDGs}{Sustainable Development Goals}
\newacronym{usv}{USV}{Unmanned Surface Vehicle}

%% file: 0-abstract.tex
As solar power continues to grow and replace traditional energy sources, the need for reliable forecasting models becomes increasingly important to ensure the stability and efficiency of the grid. However, the management of these models still needs to be improved, and new tools and technologies are required to handle the deployment and control of solar facilities. This work introduces {a novel framework named \gls{caide}, designed} for real-time monitoring, management, and forecasting of solar irradiance sensor farms. \gls{caide} is designed to manage multiple sensor farms simultaneously while improving predictive models in real-time using well-grounded \gls{ms} methodologies. The framework leverages \gls{mbse} and an \gls{iot} infrastructure to support the deployment and analysis of solar plants in dynamic environments. The system can adapt and re-train the model when given incorrect results, ensuring that forecasts remain accurate and up-to-date. Furthermore, \gls{caide} can be executed in sequential, parallel, and distributed architectures, assuring scalability. The effectiveness of \gls{caide} is demonstrated in a complex scenario composed of several solar irradiance sensor farms connected to a centralized management system. Our results show that \gls{caide} is scalable and effective in managing and forecasting solar power production while improving the accuracy of predictive models in real time. The framework has important implications for the deployment of solar plants and the future of renewable energy sources.

%% file: 1-introduction.tex

Solar energy has become increasingly important in recent years due to its potential to reduce greenhouse gas emissions and mitigate climate change. It can help to reach the promise of net zero emissions by 2050 \cite{Bouckaert2021}. Solar \gls{pv} systems are a popular and promising source of renewable energy, which is growing at the highest rate in the \gls{eu} \cite{EUSolar}. However, production can be variable and intermittent due to weather and other factors. This variability can make it challenging to integrate solar energy into the grid, as it can lead to grid instability and even blackouts. Accurate forecasting models for \gls{pv} power are necessary to address this issue. Current models use weather data and other information to predict the output of solar \gls{pv} systems in real-time, allowing grid operators to better manage the variability of solar energy and maintain grid stability. With accurate forecasting models, grid operators can anticipate changes in solar output and adjust their operations accordingly, minimizing the risk of blackouts and other disruptions \cite{Yang2014}. Furthermore, accurate forecasting models can help optimize the use of solar energy and maximize its economic benefits. For example, energy traders can use these models to make more informed decisions about buying and selling solar energy. In contrast, utility companies can use them to plan and optimize their renewable energy portfolios. Ultimately, accurate forecasting models for \gls{pv} power can help accelerate the transition to a more sustainable and resilient energy system \cite{Zhang2018}.


Most modern \gls{pv} models are based on accurate predictions of what are called \emph{base models}. They anticipate the amount of energy produced, starting with either the corresponding initial state (classic models) or with a training/inference data window (heuristic models). The current methods explored in the field can be grouped as \gls{nwp}, image-based, statistical, and \gls{ml}. They can also be classified based on their characteristics, meaning whether a method takes into account spatio-temporal features or not, is deterministic or probabilistic, considers exogenous inputs (other inputs such as physical variables) or just its data features, etc. \cite{Yang2019}. For instance, Arbizu-Barrena et al. \cite{Arbizu2017} focus on the cloud index to forecast solar irradiance with the aid of an \gls{nwp}. Ayet and Tandeo \cite{Ayet2018} aim to forecast solar irradiance based on geostationary satellite images. \gls{ml} approaches are recently gaining attention: Alzahrani et al. \cite{Alzahrani2017}, for example, employ \gls{lstm} networks for high-resolution forecasting (100Hz) in a single location, obtaining high levels of precision. The family of models published by \cite{Prado2021} comprises several \glspl{ann} for solar irradiance forecasting, where the scenario is not bound to the specific number of available sensors or their distribution, and constitute a more resilient approach since algorithms can recover from sensor failures. These features, such as flexibility and robustness, are paramount for developing a more general prediction framework and are discussed in Section~\ref{sub:nffs}.


{The utilization of \emph{base models}, which are integrated into generic software tools such as RatedPower \cite{RatedPower}, PVsyst \cite{Kumar2021}, or Helioscope \cite{Milosavljevic2022}, has been instrumental for companies to plan, design, and optimize the \gls{pv} plant engineering process, thereby maximizing profitability. These base models are adept at studying specific aspects of complex systems or predicting isolated variables within the entire system.}

\begin{figure}
  \centering
  \includegraphics[width=0.9\textwidth]{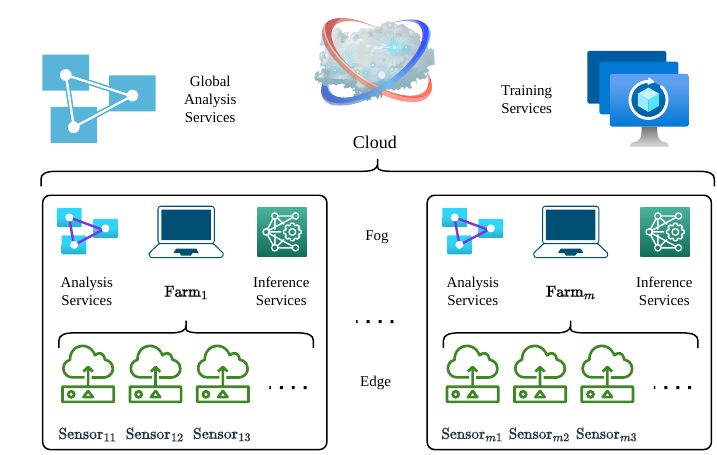}
  \caption{Big picture of the proposed integrative model.}
  \label{fig:big_picture}
\end{figure}

However, the research detailed in this paper shows that relying solely on base models is insufficient for capturing the dynamic and interconnected nature of solar irradiance sensor farms. Its vision embraces a system of systems architecture, which is referred to as an \emph{integrative model}. This model transcends the traditional base models and associated \glspl{gui} by modeling the entire ecosystem, including sensors, forecasting models, servers, domain expert actions, analysis services, and more. Success stories in other research domains, such as flood detection \cite{Basha2008}, water treatment \cite{J-RiscoMartin2023}, and healthcare \cite{Henares2022}, underscore the efficacy of such integrative approaches. Figure \ref{fig:big_picture} illustrates the conception of the integrative model proposed, which is designed to manage the complexities of deploying and operating multiple solar irradiance sensor farms, each interconnected with a centralized management system. By incorporating both real and virtual replicas of sensors, the framework facilitates a comprehensive analysis of \gls{pv} solar production possibilities and ensures the robustness and adaptability of the predictive models within a dynamic environment.

Consequently, the data generated by the sensors can also be synthetic or authentic. This complex system can be conceptualized first by using a scalable complex model, where, through a well-structured \gls{ms} methodology, all the aforementioned \emph{base models} can be easily integrated. Following an \gls{iot}-based architecture, solar irradiance is monitored at the edge layer by a set of sensors that continuously send data to the server at the fog layer. There, domain experts can analyze data, run some tests, or schedule the execution of predictive models. A cloud layer also exists, where authorities can compare different reports and make high-level decisions. The power of the cloud layer can also be used to train predictive base models, ensuring the scalability and durability of the system. 

In this paper, a new framework called \gls{caide} that enables real-time monitoring and prediction of solar energy for \gls{pv} sensor plants is introduced. This approach aims to ensure the reliability and scalability of infrastructure design and deployment while providing high-performance real-time services, such as outlier detection, complex forecasting, and training algorithms for model accuracy. This is achieved using model-driven technologies and an infrastructure based on \gls{iot}. Three main topics in the sustainable management of solar irradiance monitoring and prediction using model-driven technologies are addressed: (i) providing a robust interface to manage different solar irradiance sensor farms simultaneously, (ii) vertical scalability by modeling the entire structural pyramid from sensors to authorities, and (iii) horizontal scalability, which allows for the addition of more sensors and farms with the support of parallel and distributed simulation.

The framework is built upon a conceptual layer, using formal models and synthetic data to prove the feasibility of the entire architecture. To this end, the virtual structure and modular behavior of \gls{caide} are defined, which is capable of carrying out initial synthetic experiments through \gls{devs} \cite{Zeigler2018}, a well-known \gls{ms} formalism. Actual data can feed the framework, and virtual sensors can be replaced by physical ones following an incremental \gls{mbse} procedure. 

Overall, our main contributions are as follows:
\begin{itemize}
    \item \textbf{The \gls{caide} framework}: The \gls{caide} framework has been developed, which serves as an integrated platform for studying and analyzing solar irradiance data. By combining various models and techniques, \gls{caide} promotes collaboration and synergy among different aspects of solar energy research.
    \item \textbf{Development of a predictive solar irradiance model}: This work focuses on modeling solar irradiance within the \gls{caide} architecture. To this aim, the modeling of irradiance maps, which involves advanced techniques for spatial modeling of solar irradiance, is addressed. Additionally, the beneficial features that integrating such a model brings are explored.
    \item \textbf{Outlier detection, data analysis, and report generation}: Detection of outliers in solar irradiance data is performed using the Prophet model \cite{Toharudin2023}. While primarily designed for time series forecasting, the Prophet model also proves to be effective for outlier detection. Furthermore, an analysis subsystem within the \gls{caide} architecture has been developed, which is vital for examining the stored information and generating comprehensive reports.
\end{itemize}

This paper is organized as follows. Section \ref{sec:background} presents the foundational technologies used to design the integrative model. Section \ref{sec:architecture} offers the architecture of the framework, based on a well-known \gls{ms} formalism and able to perform parallel and distributed simulations. Section \ref{sec:services} elaborates on the main elements of the implemented predictive support and analysis subsystem. Section \ref{sec:experiments} illustrates the simulations performed to test the previously formulated hypotheses and shows the results obtained in a hybrid scenario, fed with real monitoring and synthetic data. Finally, Section \ref{sec:conclusion} draws some conclusions and introduces future research lines.

%% file: 2-background.tex
The integrative framework must be able to run scalable simulation scenarios based on the template provided in Figure \ref{fig:big_picture}. To build the integrative model, a \gls{ms} formalism named \gls{devs} \cite{Zeigler2018} has been selected. {\gls{devs} offers several distinct advantages over other \gls{ms} methods like Petri Nets or Timed Automata, particularly for the development of complex, data-driven systems like \gls{caide}. Firstly, \gls{devs} provides a clear and rigorous formalism for modeling discrete event systems, which is essential for ensuring the correctness and verifiability of the simulations. Its modular and hierarchical structure facilitates the decomposition of complex systems into manageable components, promoting reusability and maintainability. Moreover, \gls{devs} inherently supports parallel and distributed simulation, allowing for scalable and efficient execution across various computational environments, from edge devices to cloud servers. Additionally, the well-defined separation of model and simulator within the formalism enhances the flexibility to adapt to changes in system requirements or configurations. In this regard,} the \gls{devs} \gls{api} used to implement \gls{caide} has been xDEVS \cite{J-RiscoMartin2022b}. This library includes repositories for C, C++, C\#, Go, Java, Python, and Rust that provide equivalent \gls{devs} interfaces. In particular, \gls{caide} uses the xDEVS/Python module of the project. xDEVS provides support to use virtual or real-time. Additionally, it can manage sequential, parallel, or distributed simulations without modifying a single line of code in the underlying simulation model \cite{J-RiscoMartin2022}.

Around the \gls{devs} specification of \gls{caide}, some predictive, training, outlier detection, and analysis services have been built. These services work over a base predictive model from the family of \gls{dl} models published in \cite{Prado2021}. For the sake of completeness, the following foundational technologies and concepts from which \gls{caide} has been developed are introduced: the \gls{devs} formalism and xDEVS, and the required features of the \gls*{dl}-based method used.

\subsection{The \gls{devs} formalism}

The simulation framework presented in this article is based on the ground foundations of parallel \gls{devs}. 

Parallel \gls{devs} is a formal method used to model discrete event systems utilizing set theory~\cite{Zeigler2018}. It is comprised of atomic and coupled models that can communicate with other models through input ($X$) and output ($Y$) ports. Each atomic model has a state ($S$) that is associated with a time advance function $ta$, which determines the duration of the state.

Once the time assigned to the state has passed, an internal transition is triggered, and an internal transition function ($\delta_{\rm int}: S \rightarrow S$) is executed, producing a local state change ($\delta_{\rm int}(s) = s'$). At that time, the results of the model execution are spread through the output ports of the model by activating an output function ($\lambda$). 

Furthermore, external input events (received from other models) are collected in the input ports. An external transition function ($\delta_{\rm ext}: S \times e \times X \rightarrow S$) specifies how to react to those inputs, using the current state ($s$), the elapsed time since the last event ($e$) and the input value ($x$) ($\delta_{\rm ext}((s, e), x) = s'$). Parallel \gls{devs} introduces a confluent function ($\delta_{\rm con}((s, ta(s)), x) = s'$), which decides the next state in cases of collision between external and internal transitions. 

Coupled models are created by linking two or more atomic or coupled models through explicit couplings. This feature enables the use of networks of systems as components in larger coupled models, resulting in hierarchical and modular designs. 

In summary, the \gls{devs} formalism offers many advantages for analyzing and designing complex systems, including completeness, verifiability, extensibility, and maintainability. 

Once a system is described according to \gls{devs} theory, it can be easily implemented using one of the many \gls{devs} \gls{ms} engines that have come into existence in the last decades. 

Among them, xDEVS \cite{J-RiscoMartin2022} offers an excellent alternative to parallelize or distribute simulations in the Cloud, following a microservices architecture and containerization. As a result, any \gls{devs} model can be parallelized or distributed by assigning resources (threads or processes) to different transition and output functions as parallel or distributed functional programming.

\subsection{Non-Functional Features when forecasting solar irradiance}\label{sub:nffs}

As discussed in Section~\ref{sec:introduction}, anticipating the amount of solar energy that can be produced in a given area is fundamental.
In the scenario studied, two solar farms, each with a variable number and distribution of irradiance sensors are managed.
Therefore, the deployed forecasting framework must be flexible and robust concerning different sensor configurations.
These characteristics are commonly known as \textit{\glspl{nff}} \cite{Prado2021CEDI}.

In this work, a \gls*{dl}-based forecasting model from a previous work \cite{Prado2021} is incorporated into the \gls{caide} framework.
Appendix~\ref{sec:appendix} discusses the main aspects of the forecasting model.
Once trained, the model can be deployed on the corresponding fog server, allowing one to run asynchronous predictions at any moment.
The model can also be trained in the cloud layer with new data accumulated by the system, allowing one to update its weights over time when needed.  

Thanks to the forecasting process, the model gains the following \glspl{nff}:
\begin{itemize}
	\item \textbf{Flexibility}: Ability to adapt to a variable number of sensors on both the input and the output. This feature implies that if new sensors are installed in any farm or existing ones are removed, the framework can absorb these changes (see Appendix~\ref{sec:appendix}).
    \item \textbf{Robustness}: Ability to recover from missing data due to sensor failure and continue producing reliable predictions. This \gls{nff} is elaborated in Appendix~\ref{sec:appendix}.
	\item \textbf{Portability}: Ability to maintain the same forecasting architecture among different solar farms after retraining it. Therefore, the structure of the model can remain the same across farms even with different sensor arrangements, facilitating scalability and maintainability. This property is further discussed in Section~\ref{sub:model}.
\end{itemize}

These \glspl{nff} are paramount for the described scenario since they allow for the development of an independent and equivalent model for each farm, even with different numbers of sensors and spatial configurations.
Furthermore, updating the model when needed provides a tool for maintaining reasonable levels of accuracy on each farm for the foreseeable future.

%% file: 3-architecture.tex

\gls{caide}'s model divides the proposed framework for sensor farm management into the three classical IoT layers: edge, fog, and cloud. In this case, as Figure \ref{fig:big_picture} depicts, the \textit{edge} layer includes all the irradiance sensors connected to the Internet while generating data. The data generated by these sensors are sent to the next layer for further processing. The \textit{fog} layer is an intermediate layer between the edge and the cloud. This layer includes devices with computing power and storage capabilities to perform basic data processing and analysis. The fog layer is responsible for processing data in real-time and reducing the amount of data that needs to be sent to the cloud for further processing. The \textit{cloud} layer includes cloud servers and data centers that can store and process large amounts of data. The cloud layer performs complex data analytics and machine learning tasks that require significant computing power and storage capacity. 

\begin{figure}
  \centering
  \includegraphics[width=0.85\textwidth]{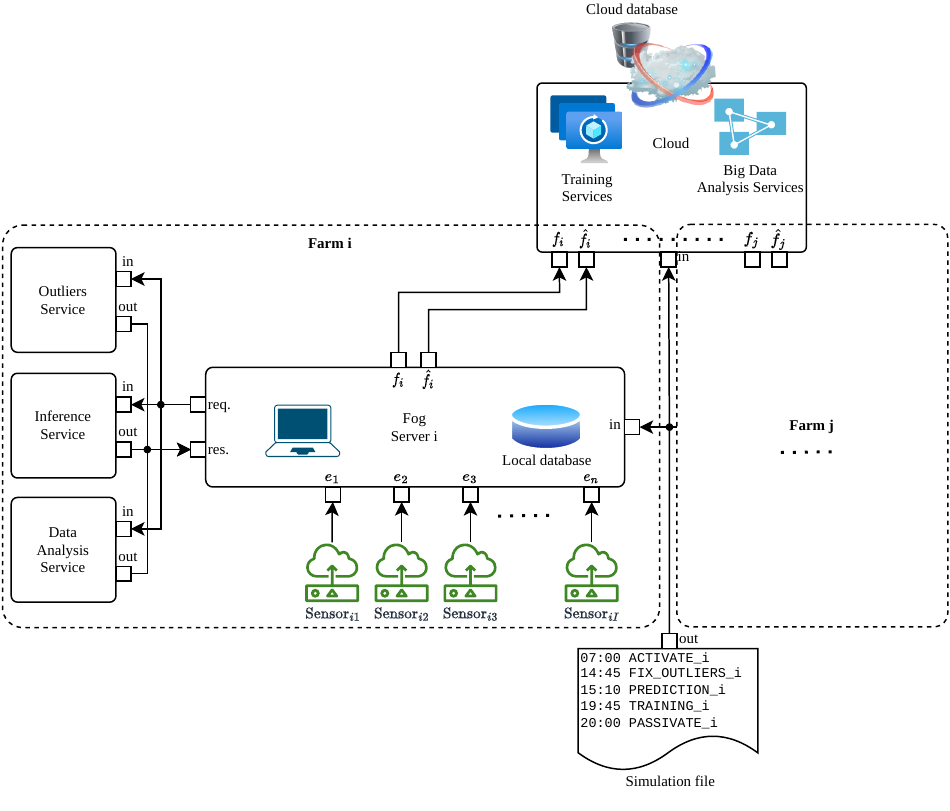}
  \caption{\gls{devs} system architecture.}
  \label{fig:devs_arch}
\end{figure}

While Figure \ref{fig:big_picture} has already illustrated the general picture of the architecture of the framework, Figure \ref{fig:devs_arch} depicts its \gls{devs} structure, which is described below. The high-level architecture of \gls{caide} is described instead of formally describing the \gls{devs} structure and behavior of all its atomic and coupled models, which is detailed inside the \gls{caide} source code\footnote{https://github.com/jlrisco/caide}. The reader is referred to Section \ref{sec:background} to better follow this section through the \gls{devs} formalism.

Figure \ref{fig:devs_arch} depicts the root coupled model. The components included in this coupled model are a simulation file, $F$ farms, and the cloud.

The \emph{simulation file} atomic model is a singular source that reads from a text file all events that are injected into the simulation process through its output port \emph{out}. Each entry in this file has a time mark indicating the virtual instant in which this event will be triggered, the associated command type, and the arguments that each command needs. As a result, this file replicates the set of external events that could happen in a real-world scenario. {The simulation file is an integral part of the discrete event simulation model. It serves as an atomic model that dictates the sequence of events and commands that drive the entire simulation process. This includes the activation and deactivation of sensors, the triggering of services such as outlier detection and forecasting, and the scheduling of model training. The simulation file essentially acts as a representation of the operator's instructions, which are crucial for the dynamic behavior of the simulation.} As the excerpt of Figure \ref{fig:devs_arch} illustrates, it always begins and ends with the triggering of the initialization and finalization of the simulation experiment (see \texttt{ACTIVATE} and \texttt{PASSIVATE} commands). 

The \emph{farm} coupled model represents a set of solar irradiance sensors located in a geographical area of the Earth and with a control station to monitor and control the facility. This coupled model has an input port, which receives the events sent by the \emph{simulation file} atomic model, and two output ports that send raw data collected by the sensors to the cloud, as well as augmented or fixed sensor data through outlier detection or data analysis services. The \emph{farm} coupled model contains several atomic models:

\begin{itemize}
\item A set of $S$ solar irradiance sensor atomic models. All are located in the edge layer of our \gls{iot} architecture. The atomic model incorporates a set of parameters to mimic the behavior of an actual sensor, like delay, minimum and maximum values, precision and noise distributions, etc. {As a result, these software components simulate the behavior of real-world sensors. They are designed to read data, either from a file containing historical or synthetic data or from real-time data streams, and then send this data to the appropriate services within the \gls{caide} framework. These services could include data processing, outlier detection, forecasting, and other analytical tasks. The atomic sensor models are not just passive data readers; they can also incorporate additional functionalities such as simulating sensor noise, delays, and failures to create a more realistic representation of sensor behavior.}
\item A set of atomic models in charge of executing services. Three services have been deployed: one to detect and fix outliers, another to perform inference of the selected predictive algorithm, and the last to perform data analysis and report. They all follow the same atomic model template. Currently, the services are executed as part of the fog layer, i.e., as local processes (to the control station or fog server) because, as will be shown in Section \ref{sec:experiments}, they do not need high computational resources. However, this can be modified at any time by adding more computing resources in the fog layer or externalizing some services to the cloud.
\item An atomic model representing the server located at the control station, named \emph{fog server}. Firstly, it receives simulation commands from the \emph{simulation file} atomic model, which tells the server when to start reading data, execute an outlier detection service, an inference, etc. Sensor data are received through the $e_i$ input ports when the simulation begins. When a data set is received, it is stored in the local database and sent to the cloud atomic model through the $f$ output port. On the other hand, when a service request is received from the \emph{simulation file}, it is propagated to the corresponding atomic model. Fixed or predicted data are stored in the local database and sent to the cloud atomic model through the $\hat{f}$ output port. 
\end{itemize}

Finally, the \emph{cloud} atomic model is located in the cloud layer. It receives all the data from the different farms (raw and estimated, i.e., fixed or predicted) and stores them in the cloud database. As in the \emph{fog server}, the cloud atomic model can run heavier services, such as performing big data analyses, including data stored on all farms, or training services to update current inference models. In any case, these actions are always commanded by the atomic model \emph{simulation file}. In this particular case, atomic models dedicated only to run services are not included because these services are always installed on virtual machines and called from a monolithic external transition function, i.e., they have a distributed architecture in nature and do not need to be encapsulated as \gls{devs} models.


The simulation file in \gls{caide} enables the handling of various events, each serving a specific purpose within the framework:

\begin{itemize}
    \item \verb|CMD_ACTIVATE_SENSORS|: This event triggers the activation of sensors at a specified timestamp, along with the destination data center, farm, and database for sensor data storage. When received by the respective sensors (refer to Figure \ref{fig:devs_arch}), they are automatically activated according to the predefined timing pattern stored in the database, considering the physical characteristics defined within the atomic models (such as delay, error, saturation statistics, etc.).
    \item \verb|CMD_PASSIVATE_SENSORS|: Similar to the previous event, this event deactivates the sensors instead of activating them. It follows the same format, including the timestamp, destination data center, farm, and database.
    \item \verb|CMD_FIX_OUTLIERS|: The Outlier Service atomic model, shown in Figure \ref{fig:devs_arch}, receives this event to perform outlier detection and repair. Along with the corresponding timestamp, it includes the destination sensor and the time interval to label the outliers. Additionally, an interpolation method (e.g., linear, quadratic, cubic, spline) can be specified for replacing outliers if desired.
    \item \verb|CMD_RUN_PREDICTION|: Invoking the prediction subsystem, this event is also managed by the Inference Service atomic model in Figure \ref{fig:devs_arch}. It includes the event timestamp, destination farm, predictive horizon, and input and output databases. As previously mentioned, the prediction is executed for all sensors within the destination farm.
    \item \verb|CMD_TRAIN_MODEL|: This event triggers the training subsystem in the cloud layer in Figure \ref{fig:devs_arch}. Along with the event timestamp, it includes the training interval and the input database. \gls{caide} provides the flexibility to externalize this service, allowing the framework to run the training service on the local host or on an external computer as a distributed simulation environment.
    \item \verb|CMD_GENERATE_REPORTS|: This service is designed to generate fog and cloud reports. The fog reports target domain experts and farm operators, providing detailed insights. On the other hand, cloud reports are intended for political authorities, offering high-level interpretations and aiding in decision-making based on coarse-grained information.
\end{itemize}

These events within the \gls{caide} framework enable efficient control and coordination of various processes, ranging from sensor activation and outlier detection to prediction generation, training, and report generation. The modular and versatile nature of these events enhances the adaptability and usability of the \gls{caide} framework for specialized users and stakeholders in the solar energy domain. \gls{caide} has been implemented using xDEVS \cite{J-RiscoMartin2022b}, a cross-platform \gls{devs} simulator. As in xDEVS, \gls{caide} can use virtual or real-time simulation. It can run sequential, parallel, and distributed simulations, or a combination of them, without modifying a single line of code in the underlying simulation model presented in Figure \ref{fig:devs_arch}. {The xDEVS simulation engine, which forms the backbone of this framework, is designed to facilitate the distribution and parallelization of models across various computing nodes. This allows for the seamless addition of multiple farms, each with its own set of sensors and control systems, without the need for rewriting simulation files when adapting to different hardware configurations. Additionally, the training phase of the predictive models is optimized for \gls{gpu} execution, providing the necessary computational power to handle extensive datasets and complex algorithms. This externalization of the training process to \glspl{gpu} ensures that the \gls{caide} framework can scale effectively during this critical phase.}

%% file: 4-services.tex
The following sections describe and elaborate on the models of the \gls{caide} architecture that are directly related to the inference and training of the predictive solar irradiance models, the detection of outliers, and the generation of data reports.

\subsection{Modeling solar irradiance}\label{sub:model}



Several physical variables can be considered when attempting to model solar irradiance using ground sensors.
\gls{ghi} is the total shortwave radiation a horizontal surface receives, which units are $\text{W}/\text{m}^2$.
Therefore, it is a good proxy for the amount of solar energy received on a surface.
Different sensors, such as pyranometers, can be deployed on solar farms to record \gls*{ghi} measurements over time.
Once a fair amount of data is collected, it can be used to train a model that forecasts short-term estimates of the energy produced in the plant.
Nevertheless, before doing so, the data must be carefully processed.
First, outliers should be removed or fixed (as will be discussed in Section~\ref{sub:outlier}).
Second, part of the missing data can be fixed using imputation methods, such as interpolation into mesh-grids (see Appendix~\ref{sec:appendix}).
Sensors usually record data with fine temporal resolutions (e.g., seconds or even milliseconds).
Such resolutions can benefit data imputation by grouping consecutive measurements into a coarser temporal granularity (e.g., minutes or hours, which can suffice depending on the application).
Furthermore, it is a common practice to standardize the data when working with neural networks, as described by Eq.~\ref{eq:stand}:
\begin{equation}\label{eq:stand}
	\widehat{x} = \frac{x - \mu(X)}{\sigma(X)}\,,\quad \forall x\in X,
\end{equation}
where $\mu(X)$ refers to the mean value of the set $X$, $\sigma(X)$ refers to the standard deviation of $X$, and $\widehat{x}$ is the standardized peer of $x$.

The inference and training services cover all these aspects when \verb|CMD_RUN_PREDICTION| or \verb|CMD_TRAIN_MODEL| commands are received, respectively.
They perform almost the same steps for every considered timestamp:
\begin{enumerate}
    \item Read the current solar irradiance values of each sensor.
    \item Standardize them based on Eq.~\ref{eq:stand}.
    \item Interpolate into mesh-grid, even if some sensor readings are missing.
    \item Stack together the last $n_x$ mesh-grids obtaining $\mathcal{X}$.
    \item Feed $\mathcal{X}$ into the model as described in Appendix~\ref{sec:appendix}, obtaining $\mathcal{Y}$.
    \item The \textbf{training service} additionally updates the model's weights based on the loss function.
    \item Reverse the interpolation on $\mathcal{Y}$ to obtain predictions for every sensor.
    \item Undo the standarization based on Eq.~\ref{eq:stand}.
    \item Store the forecasted solar irradiance values.
    \item If true values are available, the \textbf{inference service} calculates error metrics.
\end{enumerate}

The training service can be launched from scratch or for a pre-trained model.
In the latter case, the grid operator can specify the temporal interval and number of epochs used to update the model, as discussed in Section~\ref{sec:architecture}.
Assuming a pre-trained model is available, it could be the basis for forecasting solar irradiance in a similar region (size and weather-wise).
Therefore, the training service facilitates the model's portability from one farm to another.

\subsection{Outlier detection}\label{sub:outlier}

{Outlier detection is a critical feature in the application of predictive modeling for solar irradiance sensor farms due to several reasons \cite{SurveyOutliers}:}

{
\begin{itemize}
\item Solar irradiance data is susceptible to various sources of noise and errors, including sensor malfunctions, environmental obstructions (like bird droppings or foliage coverage), and atmospheric anomalies \cite{Sisodia2019}. Outliers can significantly distort the data quality, leading to inaccurate forecasts and suboptimal decision-making.
\item Predictive models, particularly those based on machine learning, are sensitive to the quality of the input data. Outliers can skew the training process, resulting in models that do not generalize well to new data \cite{Hodge2004}. By detecting and addressing outliers, it is ensured that the models learn the underlying patterns without being influenced by anomalous data points.
\item For solar farms, operational decisions such as maintenance scheduling, grid integration, and energy trading rely heavily on accurate forecasts. Outliers can lead to overestimation or underestimation of solar irradiance, which in turn can cause financial losses, grid instability, or inefficient use of resources \cite{Gandhi2024}.
\item In some cases, outliers may indicate actual extreme events, such as sudden drops in irradiance due to solar eclipses or drastic weather changes \cite{Clark2016}. While these are not errors to be corrected, detecting such outliers is crucial for implementing safety measures and adjusting operational strategies accordingly.
\item As solar farms expand and the number of sensors increases, the likelihood of encountering outliers also rises. A robust outlier detection mechanism is essential for maintaining the scalability of the system, ensuring that the predictive models can be applied across various farms with different sensor configurations and environmental conditions.
\end{itemize}
}

Outlier detection is not just a data cleaning step but a fundamental aspect of ensuring the reliability, accuracy, and robustness of predictive models in solar irradiance applications \cite{SurveyOutliers}. In this regard, CAIDE's \emph{Outliers Service} is {automatically} activated when an event of type \verb|CMD_FIX_OUTLIERS| is received. To accomplish this, the \texttt{Prophet} Python class is utilized within the atomic model's external transition function. Developed by Facebook's Core Data Science team in 2017 \cite{Toharudin2023}, Prophet is a powerful tool for time series forecasting that can also be used for outlier detection. 

Many different methods for outlier detection have been proposed such as Extended Isolation Forest \cite{EIF} or RobustSTL \cite{STL}. Any of them could be used instead of the one chosen, due to the flexibility provided by the framework, but Prophet was incorporated because it is especially suited to deal with time series with heavy periodic implications like the ones in CAIDE \cite{Prophet}. It also has simplicity, good performance, and interpretability. In this case, outliers detection typically involves a small dataset, and the Prophet training phase takes only a few seconds. This makes it ideal for repeated outlier detection. 

The Prophet class uses a decomposable time series model with three main model components: trend, seasonality, and holidays.

\begin{equation}
    y(t) = g(t) + s(t) + h(t) + \epsilon_t
\end{equation}

Prophet employs a piecewise linear trend model to capture the time series data trend. The model assumes that a series of connected linear segments can approximate the overall trend. On the other hand, Prophet models seasonal patterns using the Fourier series. It captures periodic patterns, such as daily, weekly, or yearly cycles, and can handle multiple seasonalities simultaneously. Finally, Prophet allows for the effects of holidays or special events to be included in the time series. For mathematical details on these components, the reader is referred to \cite{Vishwas2020} and \cite{Toharudin2023}.

Regarding outlier detection, Prophet does not provide a specific model for outliers. Instead, it relies on identifying points where the observed values deviate significantly from the predictions. The residual error term of the trend model $\epsilon_t$ represents the discrepancy between the observed and predicted values. Significant errors suggest potential outliers.

To perform outlier detection, one must follow several steps:

\begin{itemize}
    \item First, to \emph{prepare data}. The dataset must have two columns: a timestamp and a numeric value representing the observed values.
    \item Then, to define the Prophet Python instance and \emph{fit the model} with the previous time series data, training the Prophet model and learning the underlying patterns and trends. 
    \item Next, to \emph{generate a set of future dates} to make predictions that allow outlier detection. In this case, the same timestamp set to fit the model is used. 
    \item After, to invoke the \emph{predict method}, passing the timestamp set defined in the previous step and generating predictions based on the learned patterns.
    \item To \emph{identify outliers}. To this end, the predicted values are compared with the actual ones, calculating the difference for each data point. Outliers can be found based on those differences by choosing a threshold.
    \item Finally, to \emph{replace the outliers} using an interpolation method, for instance.
\end{itemize} 

The effectiveness of outlier detection using Prophet depends on the data quality, the chosen threshold's appropriateness, and the time series characteristics.

\subsection{Data analysis}

The analysis subsystem is a crucial component of \gls{caide}, located within the fog and cloud layers. Its primary function is to examine the information stored in the database and generate comprehensive reports. These reports include various plots that help visualize critical aspects of the simulation.

In the fog layer, a plot is produced for each sensor that contains measured data and predictions triggered by the simulation file atomic model. Additionally, it performs a comparative analysis by calculating the differences between predicted and actual values. If the error exceeds a predefined threshold, the fog server atomic model examines these disparities and signals the cloud atomic model, triggering the training service.

The report generated at the cloud layer includes a tabular representation summarizing the simulation results, including key statistical measures such as the arithmetic mean and standard deviation of measured and estimated solar irradiance, among other variables. The report{, detailed in Section \ref{sec:experiments},} also features a map depicting the sensor locations and a heat map highlighting the most productive areas.

Currently, \gls{caide} provides a set of preliminary reports to demonstrate its potential utility. However, the variety and type of report depend on the domain experts and the preferences of the decision makers. Some basic reports generated by \gls{caide} have been included in the following use case scenario.

%% file: 5-experiments.tex
\begin{figure}
    \centering
    \subfloat[Sensors in the Oahu solar measurement grid.]{%
        \includegraphics[width=0.75\textwidth]{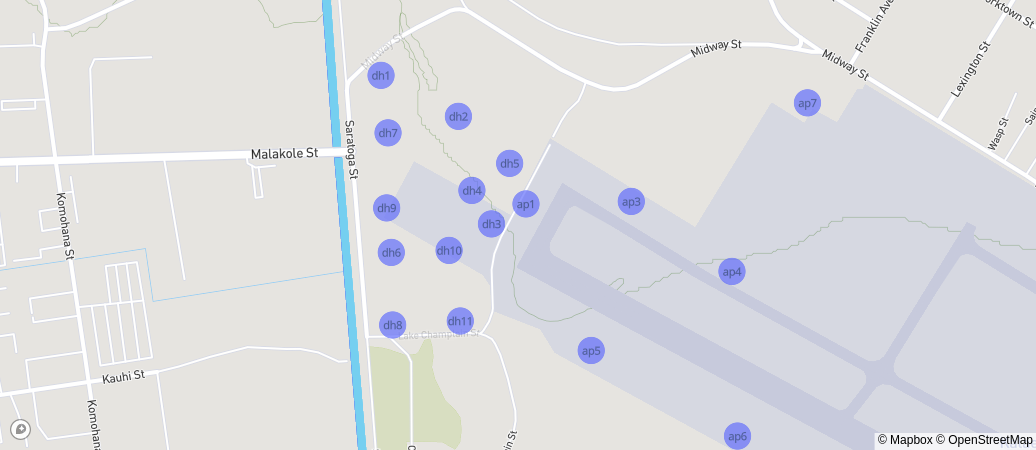}
        \label{fig:oahu-locations}
    }
    \vspace{0.2cm}
    \subfloat[Almería's synthetic grid.]{%
        \includegraphics[width=0.75\textwidth]{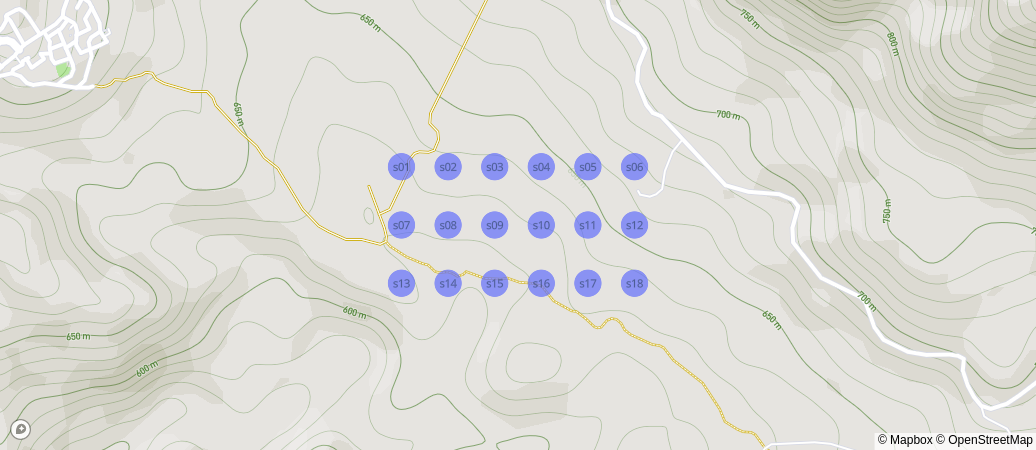}
        \label{fig:almeria-locations}
    }
    \caption{Farms location and sensors distribution.}
    \label{fig:farms-locations}
\end{figure}

In this section, a simulation using two sensor farms is conducted: one based on real monitoring data obtained from the \gls{midc}, specifically the Oahu Solar Measurement Grid \cite{oahu}, which consists of 17 sensors. These sensors are geographically distributed near the Honolulu airport, as illustrated in Figure~\ref{fig:oahu-locations}. The second farm is synthetic and comprises 18 sensors located within a localized region in Almería, Spain. The coordinates and corresponding irradiance data for these sensors have been generated using the \gls{pvgis}\footnote{https://re.jrc.ec.europa.eu/pvg\_tools}, depicted in Figure \ref{fig:almeria-locations}. Consequently, the simulated system encompasses 35 sensors distributed throughout the two farms.

An excerpt of the simulation file used to verify the correct behavior of \gls{caide} is shown below. As can be seen, both sensor farms execute the same operations: monitoring (with the activation and deactivation of the sensors), outlier detection, training, prediction, and report generation. The timestamps of the input events of Oahu and Almería are decoupled because data sources come from different moments in time. If necessary, the training service can receive an \gls{ip} address because it can be executed on another computer, different from the local host.

\begin{verbatim}
DATETIME;COMMAND;ARGUMENTS
2010-06-01 00:00:00;CMD_ACTIVATE_SENSORS;...;oahu.h5;
2010-06-27 00:00:00;CMD_TRAIN_MODEL;<ip>;...;Oahu;06-02;06-26;...
2010-06-27 01:00:00;CMD_RUN_PREDICTION;...;Oahu;06-27;...
2010-06-28 00:00:00;CMD_FIX_OUTLIERS;...;Oahu;ap1;06-27;linear;...
2010-07-01 00:00:00;CMD_PASSIVATE_SENSORS;...;Oahu;
2010-07-01 00:00:00;CMD_GENERATE_REPORTS;...;Oahu;06-01;07-01;...
2019-01-01 00:00:00;CMD_ACTIVATE_SENSORS;...;almeria.h5;
2020-03-27 00:00:00;CMD_TRAIN_MODEL;<ip>;...;Almeria;03-02;03-26;...
2020-03-27 01:00:00;CMD_RUN_PREDICTION;...;Almeria;03-27;...
2020-03-30 00:00:00;CMD_FIX_OUTLIERS;...;Almeria;s01;03-23;03-29;...
2021-01-01 00:00:00;CMD_PASSIVATE_SENSORS;...;Almeria;
\end{verbatim}

The local host has a 12th Gen Intel i7-1270P (16) @ 4.800GHz with 32 GiB RAM, Intel Alder Lake-P GPU, and the Ubuntu 22.04.2 LTS x86\_64 operating system. The training service is executed on a \gls{gcp} remote virtual machine, an Intel Xeon (4) @ 2.199GHz with 16 GiB RAM, and a NVIDIA Tesla T4, and the Ubuntu 22.04.2 LTS x86\_64 operating system.

The simulation is executed from the local host with a few lines of code as follows:

\begin{lstlisting}[language=Python, caption={\gls{caide} source code used to run the use case}]
coupled = SeveralFarms("sim-file.txt", ["Oahu", "Almeria"])
coord = Coordinator(coupled)
coord.initialize()
coord.simulate_time(INFINITY)
coord.exit()
\end{lstlisting}

Firstly, the coupled model depicted in Figure \ref{fig:devs_arch} is defined through a simulation file with all input events and the name of the two sensor farms under study. Next, the sequential \gls{devs} coordinator is instantiated. Finally, the simulation is initialized, launched, and finished. Next, all the results obtained after executing the simulation file are summarized.

\subsection{Monitoring results}

An excerpt of the simulation file used to verify monitoring sensors' correct behavior is relevant, especially for operators and domain experts in charge of the farm and with local operative decisions. \gls{caide} ensures the preservation of every record from each sensor, which is stored locally at the fog layer and the cloud layer. However, it is important to note that data is not transmitted to the cloud model in real time. Instead, it is accumulated and saved in daily packets, simulating efficient data management and transmission.

\begin{figure}
    \centering
    \includegraphics[width=0.95\textwidth]{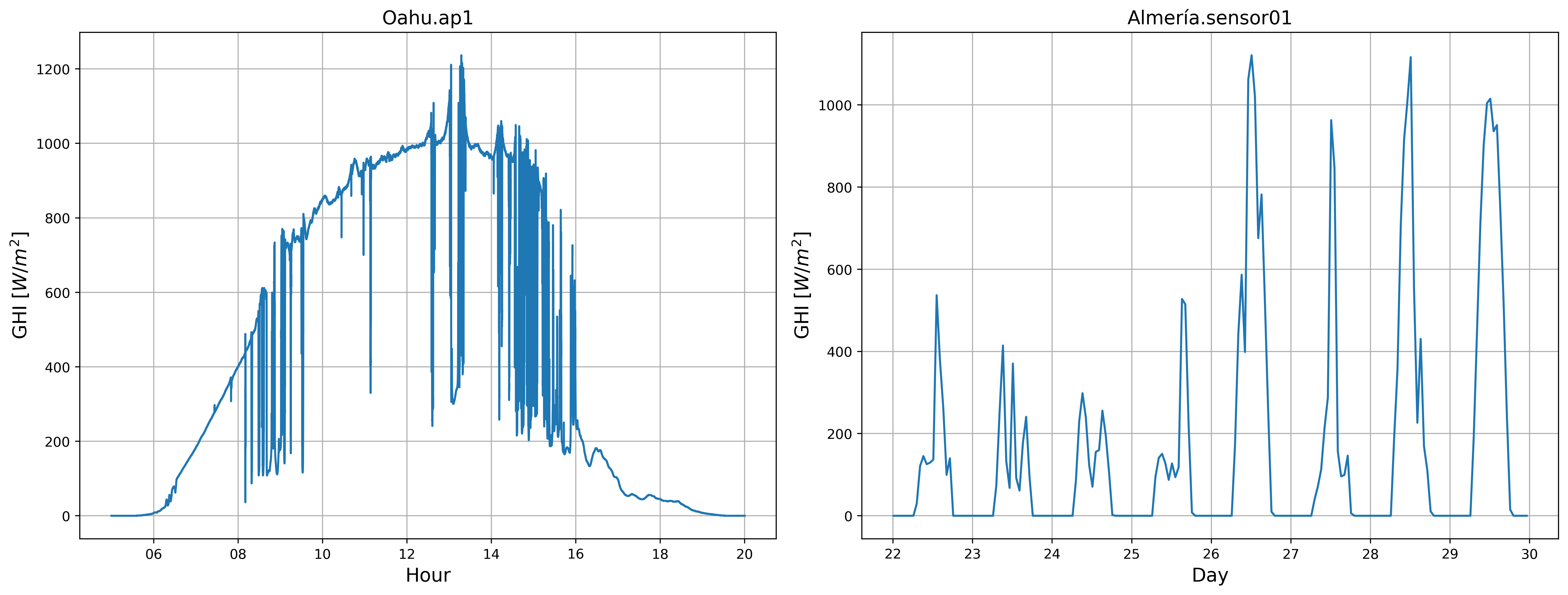}
    \caption{Representative sensor data from Oahu and Almería farms. The remaining sensors exhibit a similar profile.}
    \label{fig:representative-monitoring}
\end{figure}

{Figure \ref{fig:representative-monitoring} displays the monitoring results for two representative sensors, one from the Oahu solar farm (sensor ap1) and the other from the Almería solar farm (sensor s01). The data for sensor ap1 were recorded on June 27, 2010, a date selected due to the high solar irradiance and potential data variability. This graph shows the irradiance curves for sensor ap1, indicating the presence of some data loss and potential outliers that will be examined later in the simulation. In contrast, the data for sensor s01, which are synthetic and generated using the \gls{pvgis} tool, cover several days in March and demonstrate a more consistent pattern without apparent data loss. Despite the synthetic nature of the Almería data, outlier detection remains an essential step to ensure the robustness of the predictive models.}

{Furthermore, the monitoring phase for both farms using \gls{caide} was isolated. The simulation of the Oahu farm, with its 17 irradiation values per \emph{virtual} second, processed a total of 27,540,000 irradiation values and took approximately 167 wall-clock seconds to complete, which corresponds to 164,910 values per simulated second. The simulation of the Almería farm, with 18 irradiation values per virtual hour, processed a total of 315,792 irradiation values taking only 2 wall-clock seconds to execute. It means 157,896 values processed per second. The small difference in processing speed between the two farms is attributed to the larger size of the Oahu database, which contains more data and thus requires slightly more time for processing.}

\subsection{{Outliers detection: results}}

The outlier service allows for both the detection and correction of irradiance measurements that are not in the same range as the remaining values from the same sensor in a time window. The results of the outlier detection for both sensor farms are presented in this section. 

{In the process of outlier detection, the threshold for identifying outliers was determined by the confidence interval generated by the Prophet model, which was set to 99\%. This means that any data point falling outside the 99\% confidence interval was considered an outlier. This approach is based on the assumption that the majority of the data points will fall within the expected range of values, and those that do not are likely to be anomalies. The choice of a 99\% confidence interval is a conservative strategy aimed at minimizing the risk of false positives, where legitimate data points are incorrectly flagged as outliers. \gls{mae} and \gls{mape} were calculated to assess the model's performance and to provide a quantitative measure of the deviation of the predicted values from the actual values. The detected outliers were then corrected using linear interpolation, which is a standard method for imputing missing or anomalous values in time series data. This method assumes that the change between two data points is linear and can be used to estimate missing values within the range of known data points.}

\begin{figure}
  \centering
  \includegraphics[width=0.95\textwidth]{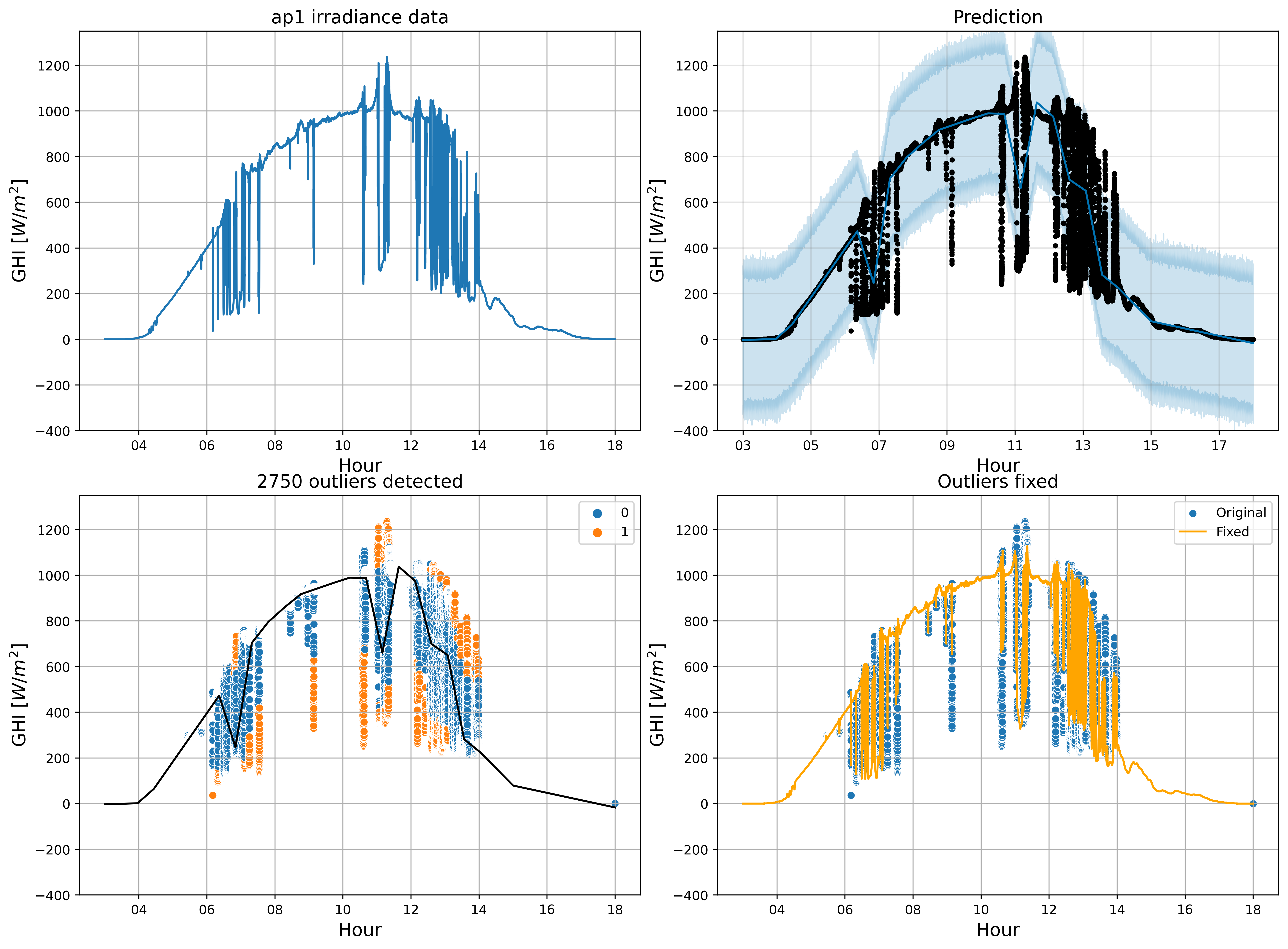}
  \caption{Oahu's sensors farm: outliers detection and treatment.}
    \label{fig:oahu-outliers}
\end{figure}

{For the Oahu sensor farm, the detection of outliers was performed on June 27, 2010, and the selected sensor was ap1. The \gls{mae} for the Prophet model was 58.35 W/m\textsuperscript{2}. Figure \ref{fig:oahu-outliers} shows the outlier detection process followed. The first subfigure illustrates the irradiance data monitored on that day. The second subfigure displays the Prophet model prediction curve, with the light blue region representing the 99\% confidence interval based on historical data, and the points representing the actual irradiance values. The third subfigure represents the detected outliers, depicted in orange as colored circles (labeled as 1). There are a total of 2750 outliers detected. The fourth subfigure displays the outliers fixed using linear interpolation.}

\begin{figure}
    \centering
    \includegraphics[width=0.95\textwidth]{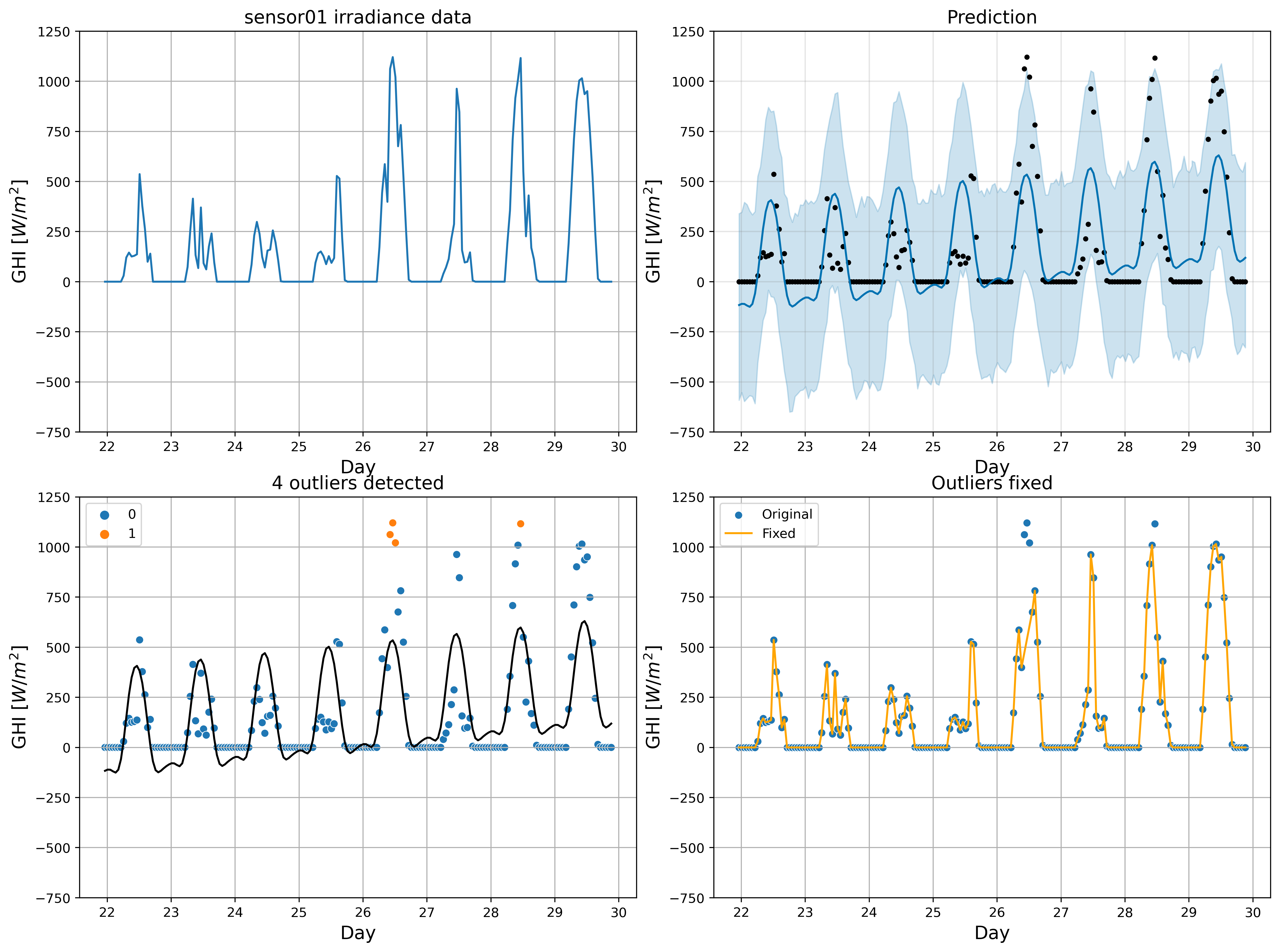}
    \caption{Almería's sensors farm: outliers detection and treatment.}
    \label{fig:almeria-outliers}
\end{figure}

{For the Almería sensor farm, since the data is synthetic and uniform, several days were selected to detect outliers, specifically from March 22, 2010, to March 29, 2010. The sensor chosen for this analysis is s01. The \gls{mae} for the Prophet model was 134.93 W/m\textsuperscript{2}, which is higher compared to the Oahu farm due to the modeling of multiple days. Figure \ref{fig:almeria-outliers} depicts the same process as in Oahu. The first subfigure shows the monitored data for those days selected during the monitoring phase. The second one shows the Prophet's prediction and the confidence interval versus real data. The third one illustrates the detected outliers, represented as orange points. Only four outliers were identified in this case. The fourth subfigure displays the outliers fixed using linear interpolation.}

As in the previous section, the outlier detection service was isolated in the simulation. It took 34 and 1 seconds to complete, respectively. As a result, the \gls{caide} outlier detection service provided by the Prophet Python class in identifying and fixing outliers in the solar irradiance sensor data from both sensor farms is very effective.

\subsection{Training service}\label{sub:training}

In this case, the \gls{dnn} is trained with data collected from the Oahu and Almería sensor farms, as detailed in Section~\ref{sub:model}.
A temporal granularity of 1 minute suffices for solar irradiance analysis.
In the case of Oahu, this implies converting the granularity from seconds to minutes by taking the average every 60 seconds.
The readings obtained with \gls{pvgis} for Almería are hourly, so the timestamps between hours are forward-filled (see Figure~\ref{fig:almeria-training}).
The \gls{dnn} is trained with accumulated data for 5, 10, 15, 20, and 25 days, respectively. The training process is conducted on a remote virtual machine specified at the beginning of this Section.

For instance, for the 25-day model, a dataset is compiled consisting of 1440 (the number of minutes in a day) multiplied by 25 (the number of days in the training period) and further multiplied by the number of sensors present in each farm (17 for Oahu and 18 for Almería). This gives a total of 612,000 irradiance data points for Oahu and 648,000 for Almería. These data points are then structured into a format suitable for the DNN model used in our study. The resulting shape of the training dataset for both farms is (35930, 10, 10, 10), where each entry represents a 10-minute sequence of data across a 10x10 grid. This grid is designed to approximate the spatial distribution of the sensors, capturing both the temporal sequence and the spatial layout of the irradiance values.

Table \ref{tab:oahu-training-performance} presents the standardized training errors (\gls{mae} and \gls{mse}) and the execution time for the Oahu and Almería sensor farms.
The \gls{mae} and \gls{mse} values show differences between the two farms.
Whereas for Almería the error decreases as more data is available, it remains stable for Oahu.
This behavior is likely caused by the more variable nature of the Oahu data, as Figure~\ref{fig:representative-monitoring} shows.
More training data and epochs would presumably revert this situation.
The training time, measured in seconds, increases {linearly} with the size of the dataset, as expected.

\begin{table}
    \centering
    \begin{tabular}{l|lll|lll|}
             & \multicolumn{3}{|c|}{Oahu} & \multicolumn{3}{|c|}{Almería} \\
        \hline
        Days & MAE & MSE & Time [s] & MAE & MSE & Time [s] \\
        \hline
        5 & 0.2612 & 0.2028 & 308 & 0.1369 & 0.1033 & 304 \\
        10 & 0.2692 & 0.2127 & 613 & 0.1395 & 0.0970 & 601 \\
        15 & 0.2486 & 0.1948 & 911 & 0.1334 & 0.0899 & 912 \\
        20 & 0.2581 & 0.2059 & 1206 & 0.1134 & 0.0640 & 1201 \\
        25 & 0.2577 & 0.2085 & 1549 & 0.1097 & 0.0559 & 1504 \\
        \hline
    \end{tabular}
    \caption{Oahu and Almería sensor farms: training errors and execution time.}
    \label{tab:oahu-training-performance}
\end{table}

As mentioned above, the \gls{caide} training service is executed on a Google Cloud Platform virtual machine in the cloud, taking advantage of the xDEVS distribution possibilities \cite{J-RiscoMartin2022}. The rest of the simulation is executed on the local host. This approach emulates a real-time deployment scenario, where the training process is externalized to the cloud to accommodate larger datasets without impacting the performance of the local infrastructure. The externalization demonstrates the scalability of the \gls{caide} framework in handling larger datasets and its ability to train the \gls{dnn} efficiently. The purpose of this work is not to extensively analyze the performance of the \gls{dnn} but to demonstrate the consistency and effectiveness of the \gls{caide} framework. 

{The \gls{devs} formalism inherently supports modularity and encapsulation, which are crucial for integrating and replacing components within a simulation framework. In the context of \gls{caide}, this means that the training algorithm, which is a core component of the predictive modeling process, can be swapped with alternative algorithms without necessitating changes to other parts of the system. This is possible because \gls{devs}-based models communicate through well-defined interfaces, characterized by their input and output ports. As long as the new training algorithm can interact with these ports by adhering to the \gls{devs} I/O conventions, it can be seamlessly integrated into the \gls{caide} framework. This modular design principle not only facilitates the maintenance and upgrading of the system but also promotes the reuse of components across different simulation scenarios, as shown in \cite{Capocchi2022}.}

The following section will discuss the forecasting results obtained by the trained \gls{dnn} models and evaluate their accuracy and reliability in predicting solar irradiance in real-time scenarios.

\subsection{Inference service}\label{sub:meth2}

A command in the simulation file can also trigger the inference service.
Once the order is executed, the model reads the $n_x$ most recent observations of each sensor.
From there, it predicts a new set of solar irradiance values for the specified forecast horizons.

Note that the number of daily predictions is restricted by the daily time window in which sensors collect data, the temporal size of the input tensor $n_x$, and the farthest prediction horizon $h$.
For example, consider data collected daily from 5:00 to 20:00 with $n_x$ = 10min and horizons $h \in \{1, 11, 31, 61\}$.
Then, the first inference will consider data from the interval [5:00, 5:09] as input to predict for instants \{5:10, 5:20, 5:40, 6:10\}.
Similarly, the last inference of the day will take as input observations from [18:50, 18:59] to forecast \gls{ghi} at the timestamps \{19:00, 19:10, 19:30, 20:00\}.
Practically, predictions can be requested up to the last input interval [19:51, 20:00].
However, the lack of observations after 20:00 implies that the error cannot be calculated in those timestamps.
The framework fixes as null observations between 20:00 and 5:00 to bypass this temporal limitation, which makes sense due to the absence of sunlight.
Furthermore, as explained above, multiple consecutive predictions can be requested to the simulator at once, as long as the input window is available.
This feature can help evaluate the model's performance on larger samples of predicted data.
Consecutive predictions could be backfed as inputs to the model for forecasting further horizons.
However, forecast accuracy will likely decline as predictions move further into the future, as seen in the subsequent experiments.


The following paragraphs present the inference (forecasting) results using the previous 25-day trained model in \gls{caide} for both the Oahu and Almería sensor farms.
Inference is carried out for a full day on each sensor farm, specifically on June 27, 2010, for Oahu and March 27, 2020, for Almería.
The inference time window covers from 05:00 to 20:00 with a temporal granularity of 1 minute, resulting in a total of 901 daily observations for every sensor (15 hours of 60 readings, plus the last one at 20:00).
The inference process in \gls{caide}, including the generation of data and the simulation of predictions, is performed on the local host.
The execution time for the inference process is 12 seconds for Oahu and 15 seconds for Almería, which is not significant enough to require distributed execution.

\begin{figure}
    \centering
    \includegraphics[width=0.9\textwidth]{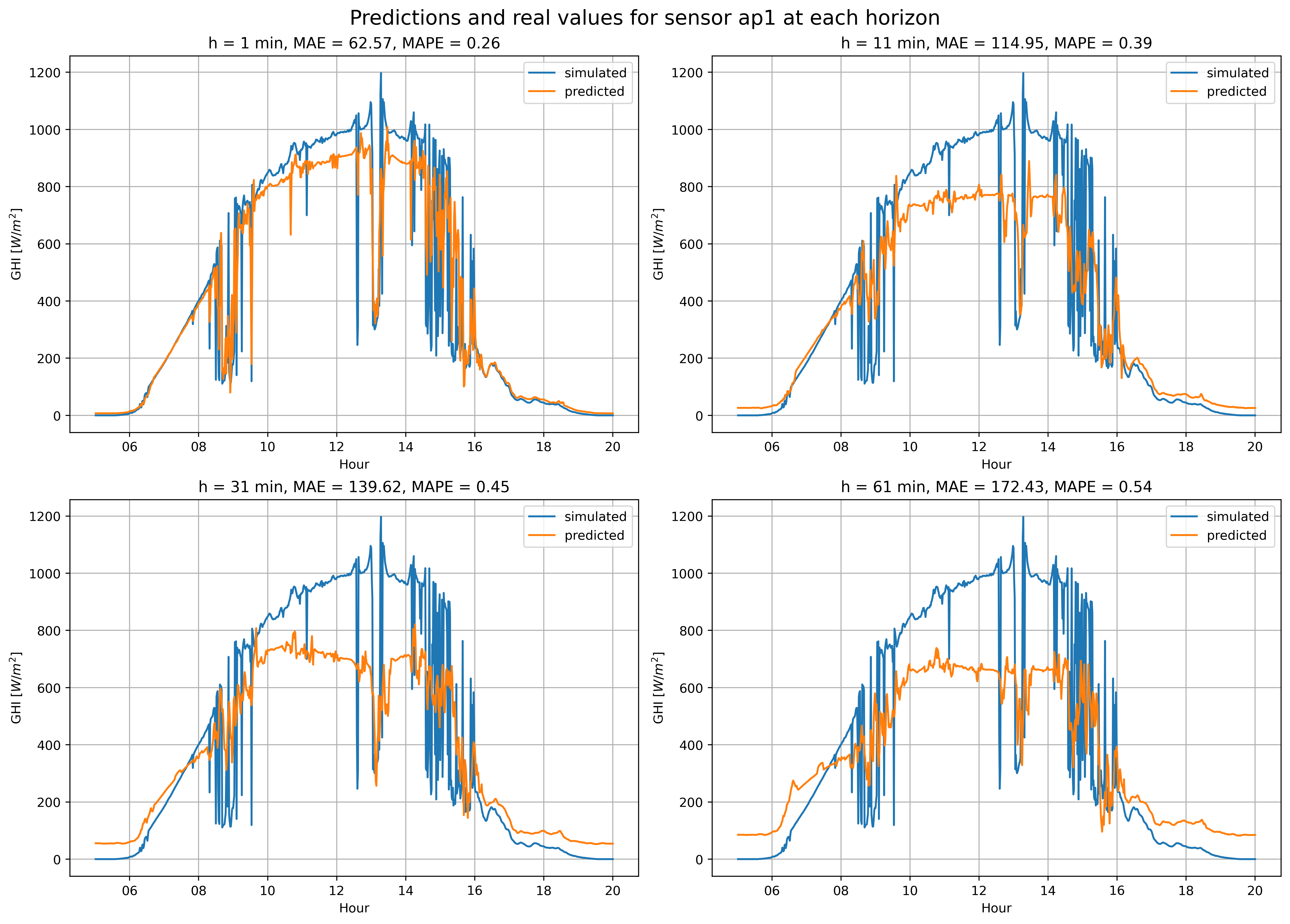}
    \caption{Predicted and simulated values at each horizon for sensor ap1 on June 27, 2010.}
    \label{fig:oahu-training}
\end{figure}

Figure \ref{fig:oahu-training} depicts the forecast results for the Oahu sensor farm for each of the four horizons considered (1, 11, 31, and 61 minutes), compared to the simulated values for the ap1 single sensor. The \glspl{mae} for each horizon are calculated to be 62.57, 114.95, 139.62, and 172.43 W/m\textsuperscript{2}, respectively. As expected, the error increases with the predictive horizon, indicating the inherent uncertainty in longer-term forecasts.

\begin{figure}
    \centering
    \includegraphics[width=0.9\textwidth]{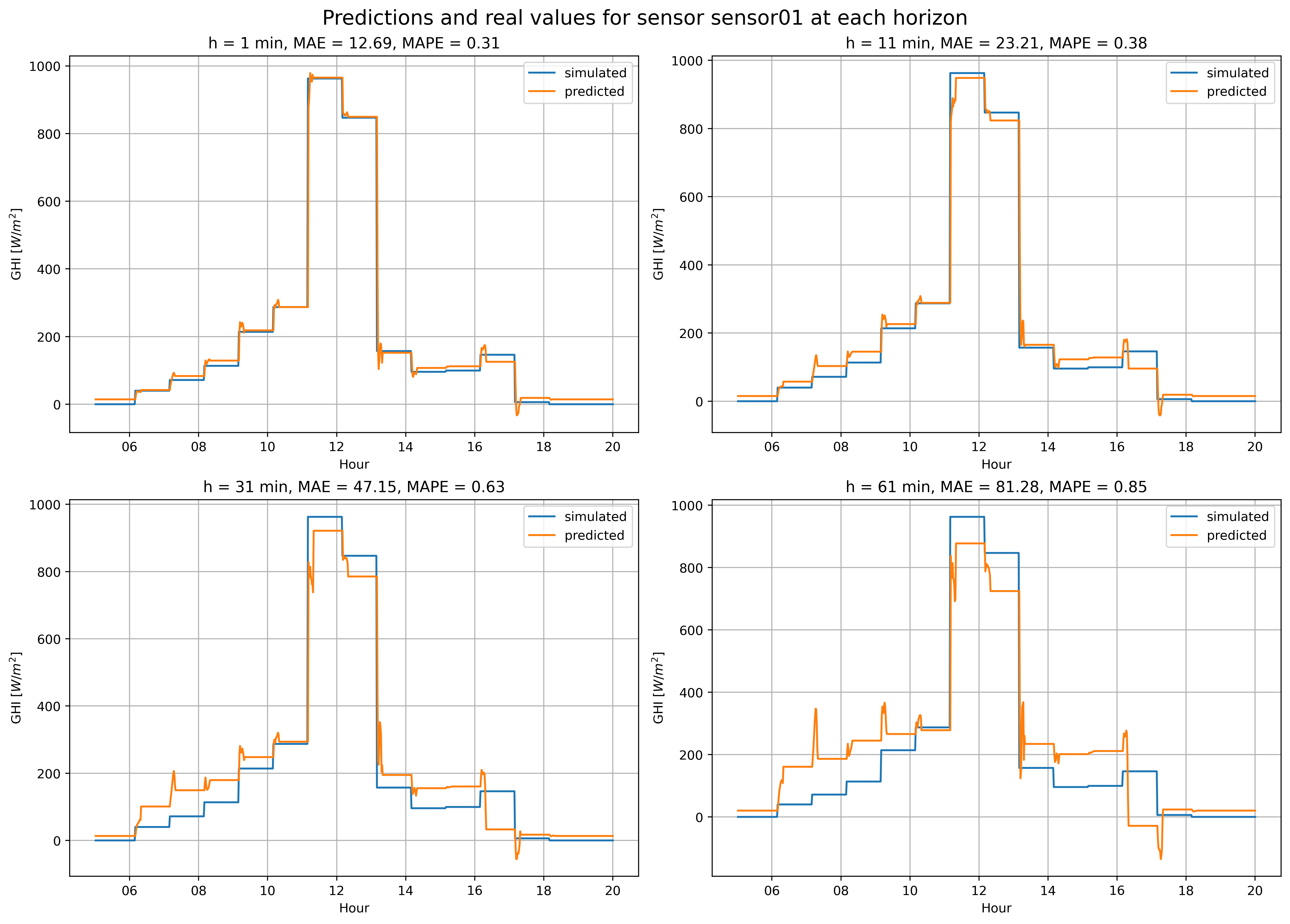}
    \caption{Predicted and simulated values at each horizon for sensor s01 on March 27, 2020.}
    \label{fig:almeria-training}
\end{figure}

Similarly, Figure \ref{fig:almeria-training} presents the forecast results for the Almería sensor farm for the same four horizons. The \glspl{mae} for each horizon are 12.69, 23.21, 47.15, and 81.28 W/m\textsuperscript{2}, respectively. These errors are relatively lower compared to the Oahu sensor farm due to the synthetic nature of the data generated using the \gls{pvgis} tool. The absence of variability and outliers in the Almería dataset contributes to the improved accuracy of the forecasts.

The results of the inference process demonstrate the capability of the 25-day trained model in \gls{caide} to provide accurate forecasts for both the Oahu and Almería sensor farms. Higher errors observed in the Oahu forecasts highlight the challenges associated with real data, including variability and outliers. {However, as highlighted above, the \gls{caide} framework is designed with a formal \gls{ms} structure that is based on the \gls{devs} formalism, which inherently allows for the straightforward substitution of predictive models. Consequently, if a predictive model needs to be replaced or updated, the process is as simple as ensuring the new model conforms to the established \gls{devs} communication interfaces. This design choice significantly reduces the complexity typically associated with integrating new predictive models, thereby streamlining the adaptation process to meet evolving requirements or to incorporate advancements in predictive methodologies.}

\subsection{Reports generation}

The ability of \gls{caide} to generate reports as HTML web pages, providing valuable insights and visualizations for the managers or decision makers of the power plant facilities is now briefly highlighted. The reports include all the figures in this section, allowing for interactive functionalities such as zooming, filtering, and data selection. The Python libraries, such as the Plotly Python graphic library, make these functionalities possible.

\begin{figure}
    \centering
    \includegraphics[width=0.95\textwidth]{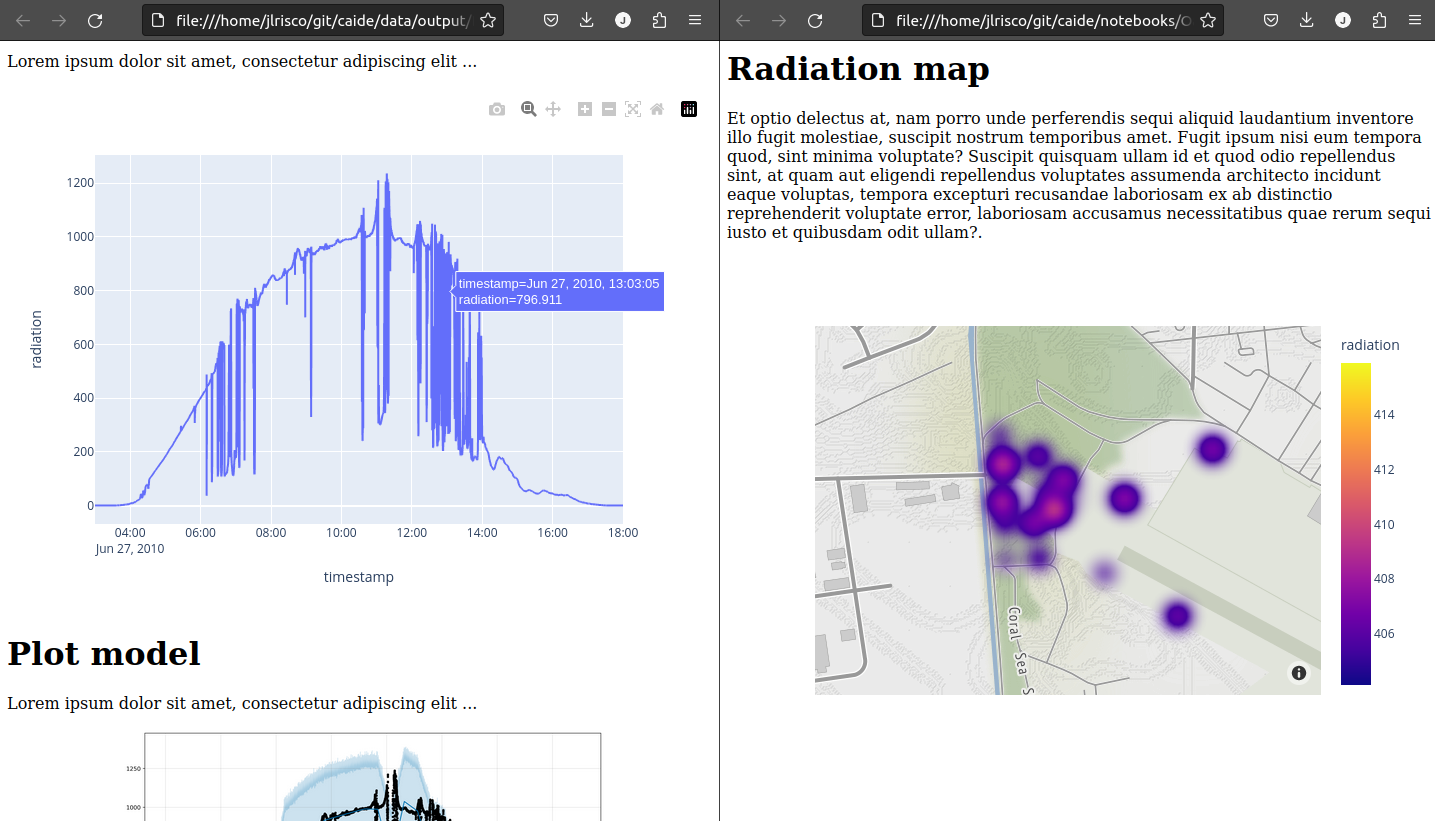}
    \caption{Oahu's preliminary generated reports.}
    \label{fig:reports}
\end{figure}

Figure \ref{fig:reports} shows a small excerpt of the outliers detection report and the heatmap generated by \gls{caide}. It is important to note that the example provided in this subsection represents a preliminary version of the reports generated by \gls{caide}. Further enhancements and refinements can be made to tailor reports to the specific needs and requirements of domain experts, managers, or decision makers.

%% file: 6-conclusion.tex

The worldwide demand for electricity is growing rapidly, surpassing the growth of renewable energy production. Consequently, there is intense anticipation for significant global expansion in renewable energy generation in the coming years. To facilitate this growth, it is imperative to develop well-organized and resilient methodologies for analyzing suitable geographic regions for power plant installations. \gls{ms} can play a crucial role in this endeavor, offering a secure and cost-effective means of obtaining an initial overview of the final deployment project. However, existing \gls{ms} approaches are not comprehensive enough to address this specific challenge in an integrative way.

In this study, \gls{caide} has been presented. This innovative and integrative framework facilitates real-time monitoring of solar irradiance sensor farms and enables decision-making regarding the advancement of cutting-edge predictive models. \gls{caide} not only detects outlier values and performs missing data estimation, but also incorporates new functionalities, such as training the predictive model as an external service executed on a remote virtual machine and generating HTML reports to present simulation results. This framework is built upon the principles of \gls{mbse} and utilizes the \gls{devs} \gls{ms} formalism. Leveraging the \gls{iot} paradigm, \gls{caide} offers a scalable and reliable infrastructure that supports incremental design and efficient management of multiple farms. Additionally, the framework provides different resolution views tailored to domain experts at the fog layer and authorities at the cloud layer, aligning with the terminology of the IoT domain.

Future work includes improving the analysis subsystem, expanding its capabilities by incorporating additional types of plots in the reports, such as time-series graphs or scatter plots, to capture different aspects of the data. Advanced statistical analysis, such as correlation analysis and regression models, could provide deeper insights into variable relationships. Including trend and spatial analysis in the reports would provide information on patterns over time and geographical influences on simulation outcomes.

%% file: 7-appendix.tex
\begin{figure}
    \centering
    \includegraphics[width=0.95\textwidth]{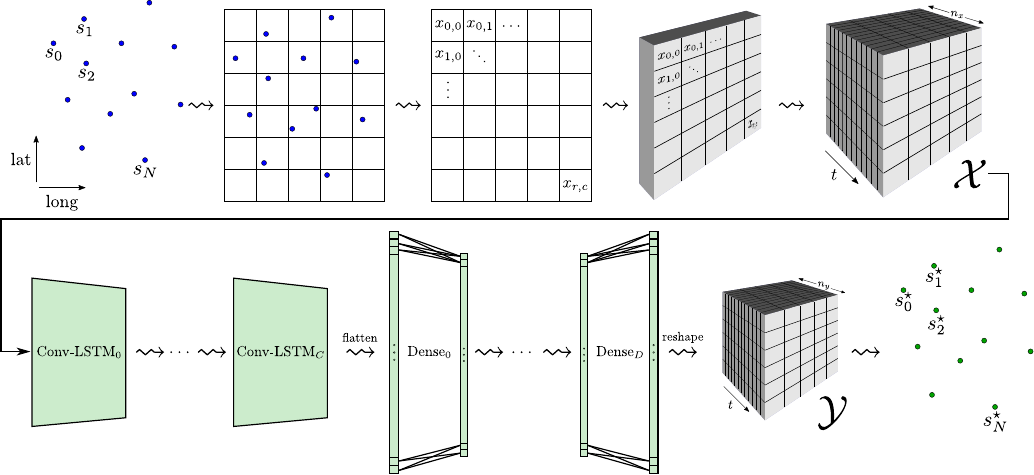}
    \caption{Diagram that shows the \gls{dl}-based model for solar irradiance forecasting from end to end.}
    \label{fig:model}
\end{figure}

As mentioned earlier, this article integrates a \gls{dl}-based model into the \gls{caide} framework, which was published in a previous work \citep{Prado2021}.
This appendix summarizes the main aspects of the forecasting model.
The step-by-step forecasting process is depicted in Figure~\ref{fig:model} and is as follows:
\begin{enumerate}
    \item The ground sensors record $N$ solar irradiance observations at the current timestamp, which are arranged arbitrarily.
    \item Afterward, these data are transformed into a two-dimensional mesh-grid using nearest-neighbor interpolation.
    \item This mesh-grid is stacked with previous ones, yielding a three-dimensional tensor.
          This structure implicitly encapsulates the temporal and spatial aspects of solar irradiance.
    \item Next, the last $n_x$ mesh-grids (denoted as $\mathcal{X}$ in Figure~\ref{fig:model}) are fed into the \gls{dnn} model.
    \item The model starts with $C$ \gls*{convLstm} layers, which use convolutional structures both in the input and in the recurrent transitions.
          Therefore, they can grasp both spatial and temporal features.
    \item Then, their output is flattened and fed into $D$ fully-connected layers.
    \item The output tensor $\mathcal{Y}$ is obtained using a reshape operation and has the same spatio-temporal structure as the input.
          The difference is that the temporal dimension encompasses the $n_y$ forecast horizons.
    \item Finally, the interpolation process is reversed.
          This step yields the $n_y \cdot N$ predicted irradiance values corresponding to each horizon and sensor location.
\end{enumerate}

The interpolation of individual sensor observations into the mesh-grid is the crucial step that equips the forecasting framework with the \glspl{nff} discussed in Section~\ref{sec:background}.
In this manner, the model effectively forgets about the specific sensors and their locations, favoring a broader view of solar irradiance maps over the region of interest.
In practice, the selection of spatial resolution and mesh-grid position must be based on the number and distribution of the sensors.
Previous experiments suggest that for the interpolation to work effectively, the sensors must cover at least 20\% of the mesh-grid pixels and be well-distributed.

Using mesh-grids favors flexibility and robustness, as seen in \cite{Prado2021}.
Robustness refers to the ability of the model to recover missing sensor observations.
Experiments from the previous work indicated that the forecasting error increases by only 10\%, even if one in four sensors fails.
Furthermore, the error worsens by 25\% or less when the number of failing sensors is increased to half of the available ones.
Therefore, these simulations indicate that the model can produce reliable predictions even under such complications.
Flexibility, as defined in Section~\ref{sec:background}, is accomplished thanks to solar irradiance mesh-grids.
When new sensors are added to the network, they can improve the mesh-grid's reliability during the interpolation step.
The only restriction is that the new locations must fall within the area covered by the grid.
Removing sensors can be done without redesigning the model as the shape of the mesh-grid remains unchanged.

%% file: main.bbl
\begin{thebibliography}{}

\bibitem [\protect \citeauthoryear {%
Alzahrani%
, Shamsi%
, Dagli%
\BCBL {}\ \BBA {} Ferdowsi%
}{%
Alzahrani%
\ \protect \BOthers {.}}{%
{\protect \APACyear {2017}}%
}]{%
Alzahrani2017}
\APACinsertmetastar {%
Alzahrani2017}%
\begin{APACrefauthors}%
Alzahrani, A.%
, Shamsi, P.%
, Dagli, C.%
\BCBL {}\ \BBA {} Ferdowsi, M.%
\end{APACrefauthors}%
\unskip\
\newblock
\APACrefYearMonthDay{2017}{}{}.
\newblock
{\BBOQ}\APACrefatitle {Solar irradiance forecasting using deep neural networks}
  {Solar irradiance forecasting using deep neural networks}.{\BBCQ}
\newblock
\APACjournalVolNumPages{Procedia Computer Science}{114}{}{304--313}.
\PrintBackRefs{\CurrentBib}

\bibitem [\protect \citeauthoryear {%
Arbizu-Barrena%
, Ruiz-Arias%
, Rodr{\'\i}guez-Ben{\'\i}tez%
, Pozo-V{\'a}zquez%
\BCBL {}\ \BBA {} Tovar-Pescador%
}{%
Arbizu-Barrena%
\ \protect \BOthers {.}}{%
{\protect \APACyear {2017}}%
}]{%
Arbizu2017}
\APACinsertmetastar {%
Arbizu2017}%
\begin{APACrefauthors}%
Arbizu-Barrena, C.%
, Ruiz-Arias, J\BPBI A.%
, Rodr{\'\i}guez-Ben{\'\i}tez, F\BPBI J.%
, Pozo-V{\'a}zquez, D.%
\BCBL {}\ \BBA {} Tovar-Pescador, J.%
\end{APACrefauthors}%
\unskip\
\newblock
\APACrefYearMonthDay{2017}{}{}.
\newblock
{\BBOQ}\APACrefatitle {Short-term solar radiation forecasting by advecting and
  diffusing MSG cloud index} {Short-term solar radiation forecasting by
  advecting and diffusing msg cloud index}.{\BBCQ}
\newblock
\APACjournalVolNumPages{Solar Energy}{155}{}{1092--1103}.
\PrintBackRefs{\CurrentBib}

\bibitem [\protect \citeauthoryear {%
Ayet%
\ \BBA {} Tandeo%
}{%
Ayet%
\ \BBA {} Tandeo%
}{%
{\protect \APACyear {2018}}%
}]{%
Ayet2018}
\APACinsertmetastar {%
Ayet2018}%
\begin{APACrefauthors}%
Ayet, A.%
\BCBT {}\ \BBA {} Tandeo, P.%
\end{APACrefauthors}%
\unskip\
\newblock
\APACrefYearMonthDay{2018}{}{}.
\newblock
{\BBOQ}\APACrefatitle {Nowcasting solar irradiance using an analog method and
  geostationary satellite images} {Nowcasting solar irradiance using an analog
  method and geostationary satellite images}.{\BBCQ}
\newblock
\APACjournalVolNumPages{Solar Energy}{164}{}{301--315}.
\PrintBackRefs{\CurrentBib}

\bibitem [\protect \citeauthoryear {%
Basha%
, Ravela%
\BCBL {}\ \BBA {} Rus%
}{%
Basha%
\ \protect \BOthers {.}}{%
{\protect \APACyear {2008}}%
}]{%
Basha2008}
\APACinsertmetastar {%
Basha2008}%
\begin{APACrefauthors}%
Basha, E\BPBI A.%
, Ravela, S.%
\BCBL {}\ \BBA {} Rus, D.%
\end{APACrefauthors}%
\unskip\
\newblock
\APACrefYearMonthDay{2008}{}{}.
\newblock
{\BBOQ}\APACrefatitle {Model-Based Monitoring for Early Warning Flood
  Detection} {Model-based monitoring for early warning flood detection}.{\BBCQ}
\newblock
\BIn{} \APACrefbtitle {Proceedings of the 6th ACM Conference on Embedded
  Network Sensor Systems} {Proceedings of the 6th acm conference on embedded
  network sensor systems}\ (\BPG~295–308).
\newblock
\APACaddressPublisher{New York, NY, USA}{Association for Computing Machinery}.
\newblock
\begin{APACrefURL} \url{https://doi.org/10.1145/1460412.1460442}
  \end{APACrefURL}
\newblock
\begin{APACrefDOI} \doi{10.1145/1460412.1460442} \end{APACrefDOI}
\PrintBackRefs{\CurrentBib}

\bibitem [\protect \citeauthoryear {%
Bouckaert%
\ \protect \BOthers {.}}{%
Bouckaert%
\ \protect \BOthers {.}}{%
{\protect \APACyear {2021}}%
}]{%
Bouckaert2021}
\APACinsertmetastar {%
Bouckaert2021}%
\begin{APACrefauthors}%
Bouckaert, S.%
, Pales, A\BPBI F.%
, McGlade, C.%
, Remme, U.%
, Wanner, B.%
, Varro, L.%
\BDBL {}Spencer, T.%
\end{APACrefauthors}%
\unskip\
\newblock
\APACrefYearMonthDay{2021}{{\APACmonth{06}}}{}.
\newblock
\APACrefbtitle {Net Zero by 2050: A Roadmap for the Global Energy Sector.} {Net
  zero by 2050: A roadmap for the global energy sector.}
\newblock
\APAChowpublished {https://trid.trb.org/view/1856381}.
\PrintBackRefs{\CurrentBib}

\bibitem [\protect \citeauthoryear {%
Boukerche%
, Zheng%
\BCBL {}\ \BBA {} Alfandi%
}{%
Boukerche%
\ \protect \BOthers {.}}{%
{\protect \APACyear {2020}}%
}]{%
SurveyOutliers}
\APACinsertmetastar {%
SurveyOutliers}%
\begin{APACrefauthors}%
Boukerche, A.%
, Zheng, L.%
\BCBL {}\ \BBA {} Alfandi, O.%
\end{APACrefauthors}%
\unskip\
\newblock
\APACrefYearMonthDay{2020}{}{}.
\newblock
{\BBOQ}\APACrefatitle {Outlier Detection: Methods, Models, and Classification}
  {Outlier detection: Methods, models, and classification}.{\BBCQ}
\newblock
\APACjournalVolNumPages{ACM Computing Surveys}{}{}{}.
\newblock
\begin{APACrefDOI} \doi{10.1145/3381028} \end{APACrefDOI}
\PrintBackRefs{\CurrentBib}

\bibitem [\protect \citeauthoryear {%
Capocchi%
\ \BBA {} Santucci%
}{%
Capocchi%
\ \BBA {} Santucci%
}{%
{\protect \APACyear {2022}}%
}]{%
Capocchi2022}
\APACinsertmetastar {%
Capocchi2022}%
\begin{APACrefauthors}%
Capocchi, L.%
\BCBT {}\ \BBA {} Santucci, J\BHBI F.%
\end{APACrefauthors}%
\unskip\
\newblock
\APACrefYearMonthDay{2022}{}{}.
\newblock
{\BBOQ}\APACrefatitle {Discrete event modeling and simulation for reinforcement
  learning system design} {Discrete event modeling and simulation for
  reinforcement learning system design}.{\BBCQ}
\newblock
\APACjournalVolNumPages{Information}{13}{3}{121}.
\PrintBackRefs{\CurrentBib}

\bibitem [\protect \citeauthoryear {%
Clark%
}{%
Clark%
}{%
{\protect \APACyear {2016}}%
}]{%
Clark2016}
\APACinsertmetastar {%
Clark2016}%
\begin{APACrefauthors}%
Clark, M\BPBI R.%
\end{APACrefauthors}%
\unskip\
\newblock
\APACrefYearMonthDay{2016}{}{}.
\newblock
{\BBOQ}\APACrefatitle {On the variability of near-surface screen temperature
  anomalies in the 20 March 2015 solar eclipse} {On the variability of
  near-surface screen temperature anomalies in the 20 march 2015 solar
  eclipse}.{\BBCQ}
\newblock
\APACjournalVolNumPages{Philosophical Transactions of the Royal Society A:
  Mathematical, Physical and Engineering Sciences}{374}{2077}{20150213}.
\PrintBackRefs{\CurrentBib}

\bibitem [\protect \citeauthoryear {%
\APACcitebtitle {{EU} {S}olar {E}nergy {S}trategy}}{%
\APACcitebtitle {{EU} {S}olar {E}nergy {S}trategy}}{%
{\protect \APACyear {2022}}%
}]{%
EUSolar}
\APACinsertmetastar {%
EUSolar}%
\APACrefbtitle {{EU} {S}olar {E}nergy {S}trategy} {{EU} {S}olar {E}nergy
  {S}trategy}\ \APACbVolEdTR{}{\BTR{}}.
\newblock
\APACrefYearMonthDay{2022}{}{}.
\newblock
\APACaddressInstitution{European Commission, 1049 Bruxelles/Brussel,
  Belgium}{European Commission}.
\newblock
\APAChowpublished
  {https://energy.ec.europa.eu/topics/renewable-energy/solar-energy\_en}.
\PrintBackRefs{\CurrentBib}

\bibitem [\protect \citeauthoryear {%
Gandhi%
\ \protect \BOthers {.}}{%
Gandhi%
\ \protect \BOthers {.}}{%
{\protect \APACyear {2024}}%
}]{%
Gandhi2024}
\APACinsertmetastar {%
Gandhi2024}%
\begin{APACrefauthors}%
Gandhi, O.%
, Zhang, W.%
, Kumar, D\BPBI S.%
, Rodr{\'\i}guez-Gallegos, C\BPBI D.%
, Yagli, G\BPBI M.%
, Yang, D.%
\BDBL {}Srinivasan, D.%
\end{APACrefauthors}%
\unskip\
\newblock
\APACrefYearMonthDay{2024}{}{}.
\newblock
{\BBOQ}\APACrefatitle {The value of solar forecasts and the cost of their
  errors: A review} {The value of solar forecasts and the cost of their errors:
  A review}.{\BBCQ}
\newblock
\APACjournalVolNumPages{Renewable and Sustainable Energy
  Reviews}{189}{}{113915}.
\PrintBackRefs{\CurrentBib}

\bibitem [\protect \citeauthoryear {%
Hariri%
, Kind%
\BCBL {}\ \BBA {} Brunner%
}{%
Hariri%
\ \protect \BOthers {.}}{%
{\protect \APACyear {2021}}%
}]{%
EIF}
\APACinsertmetastar {%
EIF}%
\begin{APACrefauthors}%
Hariri, S.%
, Kind, M\BPBI C.%
\BCBL {}\ \BBA {} Brunner, R\BPBI J.%
\end{APACrefauthors}%
\unskip\
\newblock
\APACrefYearMonthDay{2021}{}{}.
\newblock
{\BBOQ}\APACrefatitle {Extended Isolation Forest} {Extended isolation
  forest}.{\BBCQ}
\newblock
\APACjournalVolNumPages{IEEE Transactions on Knowledge and Data
  Engineering}{33}{4}{1479-1489}.
\newblock
\begin{APACrefDOI} \doi{10.1109/TKDE.2019.2947676} \end{APACrefDOI}
\PrintBackRefs{\CurrentBib}

\bibitem [\protect \citeauthoryear {%
Henares%
, Risco-Mart{\'\i}n%
, Ayala%
\BCBL {}\ \BBA {} Hermida%
}{%
Henares%
\ \protect \BOthers {.}}{%
{\protect \APACyear {2022}}%
}]{%
Henares2022}
\APACinsertmetastar {%
Henares2022}%
\begin{APACrefauthors}%
Henares, K.%
, Risco-Mart{\'\i}n, J\BPBI L.%
, Ayala, J\BPBI L.%
\BCBL {}\ \BBA {} Hermida, R.%
\end{APACrefauthors}%
\unskip\
\newblock
\APACrefYearMonthDay{2022}{}{}.
\newblock
{\BBOQ}\APACrefatitle {Efficient micro data centres deployment for mobile
  healthcare monitoring systems in IoT urban scenarios} {Efficient micro data
  centres deployment for mobile healthcare monitoring systems in iot urban
  scenarios}.{\BBCQ}
\newblock
\APACjournalVolNumPages{Journal of Simulation}{}{}{1--15}.
\PrintBackRefs{\CurrentBib}

\bibitem [\protect \citeauthoryear {%
Hodge%
\ \BBA {} Austin%
}{%
Hodge%
\ \BBA {} Austin%
}{%
{\protect \APACyear {2004}}%
}]{%
Hodge2004}
\APACinsertmetastar {%
Hodge2004}%
\begin{APACrefauthors}%
Hodge, V.%
\BCBT {}\ \BBA {} Austin, J.%
\end{APACrefauthors}%
\unskip\
\newblock
\APACrefYearMonthDay{2004}{}{}.
\newblock
{\BBOQ}\APACrefatitle {A survey of outlier detection methodologies} {A survey
  of outlier detection methodologies}.{\BBCQ}
\newblock
\APACjournalVolNumPages{Artificial intelligence review}{22}{}{85--126}.
\PrintBackRefs{\CurrentBib}

\bibitem [\protect \citeauthoryear {%
Kumar%
, Rajoria%
, Sharma%
\BCBL {}\ \BBA {} Suhag%
}{%
Kumar%
\ \protect \BOthers {.}}{%
{\protect \APACyear {2021}}%
}]{%
Kumar2021}
\APACinsertmetastar {%
Kumar2021}%
\begin{APACrefauthors}%
Kumar, R.%
, Rajoria, C.%
, Sharma, A.%
\BCBL {}\ \BBA {} Suhag, S.%
\end{APACrefauthors}%
\unskip\
\newblock
\APACrefYearMonthDay{2021}{}{}.
\newblock
{\BBOQ}\APACrefatitle {Design and simulation of standalone solar {PV} system
  using {PV}syst software: a case study} {Design and simulation of standalone
  solar {PV} system using {PV}syst software: a case study}.{\BBCQ}
\newblock
\APACjournalVolNumPages{Materials Today: Proceedings}{46}{}{5322--5328}.
\PrintBackRefs{\CurrentBib}

\bibitem [\protect \citeauthoryear {%
Milosavljevi{\'c}%
, Kevki{\'c}%
\BCBL {}\ \BBA {} Jovanovi{\'c}%
}{%
Milosavljevi{\'c}%
\ \protect \BOthers {.}}{%
{\protect \APACyear {2022}}%
}]{%
Milosavljevic2022}
\APACinsertmetastar {%
Milosavljevic2022}%
\begin{APACrefauthors}%
Milosavljevi{\'c}, D\BPBI D.%
, Kevki{\'c}, T\BPBI S.%
\BCBL {}\ \BBA {} Jovanovi{\'c}, S\BPBI J.%
\end{APACrefauthors}%
\unskip\
\newblock
\APACrefYearMonthDay{2022}{}{}.
\newblock
{\BBOQ}\APACrefatitle {Review and validation of photovoltaic solar simulation
  tools/software based on case study} {Review and validation of photovoltaic
  solar simulation tools/software based on case study}.{\BBCQ}
\newblock
\APACjournalVolNumPages{Open Physics}{20}{1}{431--451}.
\PrintBackRefs{\CurrentBib}

\bibitem [\protect \citeauthoryear {%
Prado-Rujas%
, Garc{\'\i}a-Dopico%
, Serrano%
\BCBL {}\ \BBA {} P{\'e}rez%
}{%
Prado-Rujas%
, Garc{\'\i}a-Dopico%
\BCBL {}\ \protect \BOthers {.}}{%
{\protect \APACyear {2021}}%
}]{%
Prado2021}
\APACinsertmetastar {%
Prado2021}%
\begin{APACrefauthors}%
Prado-Rujas, I\BHBI I.%
, Garc{\'\i}a-Dopico, A.%
, Serrano, E.%
\BCBL {}\ \BBA {} P{\'e}rez, M\BPBI S.%
\end{APACrefauthors}%
\unskip\
\newblock
\APACrefYearMonthDay{2021}{}{}.
\newblock
{\BBOQ}\APACrefatitle {A flexible and robust deep learning-based system for
  solar irradiance forecasting} {A flexible and robust deep learning-based
  system for solar irradiance forecasting}.{\BBCQ}
\newblock
\APACjournalVolNumPages{IEEE Access}{9}{}{12348--12361}.
\newblock
\begin{APACrefDOI} \doi{10.1109/ACCESS.2021.3051839} \end{APACrefDOI}
\PrintBackRefs{\CurrentBib}

\bibitem [\protect \citeauthoryear {%
Prado-Rujas%
, Serrano%
, García-Dopico%
, Córdoba%
\BCBL {}\ \BBA {} Pérez%
}{%
Prado-Rujas%
, Serrano%
\BCBL {}\ \protect \BOthers {.}}{%
{\protect \APACyear {2021}}%
}]{%
Prado2021CEDI}
\APACinsertmetastar {%
Prado2021CEDI}%
\begin{APACrefauthors}%
Prado-Rujas, I\BHBI I.%
, Serrano, E.%
, García-Dopico, A.%
, Córdoba, M\BPBI L.%
\BCBL {}\ \BBA {} Pérez, M\BPBI S.%
\end{APACrefauthors}%
\unskip\
\newblock
\APACrefYearMonthDay{2021}{{\APACmonth{09}}}{}.
\newblock
{\BBOQ}\APACrefatitle {Predicción espacio-temporal: más allá del
  error/precisión} {Predicción espacio-temporal: más allá del
  error/precisión}.{\BBCQ}
\newblock
\BIn{} \APACrefbtitle {Congreso Español de Informática 2021 (CEDI'21).}
  {Congreso español de informática 2021 (cedi'21).}
\newblock
\APACaddressPublisher{Málaga, Spain}{}.
\newblock
\begin{APACrefDOI} \doi{10.5281/zenodo.5530048} \end{APACrefDOI}
\PrintBackRefs{\CurrentBib}

\bibitem [\protect \citeauthoryear {%
\APACcitebtitle {Rated{P}ower software tool}}{%
\APACcitebtitle {Rated{P}ower software tool}}{%
{\protect \APACyear {2023}}%
}]{%
RatedPower}
\APACinsertmetastar {%
RatedPower}%
\APACrefbtitle {Rated{P}ower software tool.} {Rated{P}ower software tool.}
\newblock
\APACrefYearMonthDay{2023}{}{}.
\newblock
\APAChowpublished {https://ratedpower.com}.
\newblock
\begin{APACrefURL} \url{https://ratedpower.com} \end{APACrefURL}
\PrintBackRefs{\CurrentBib}

\bibitem [\protect \citeauthoryear {%
Risco-Martín%
\ \protect \BOthers {.}}{%
Risco-Martín%
\ \protect \BOthers {.}}{%
{\protect \APACyear {2023}}%
}]{%
J-RiscoMartin2023}
\APACinsertmetastar {%
J-RiscoMartin2023}%
\begin{APACrefauthors}%
Risco-Martín, J\BPBI L.%
, Esteban, S.%
, Chacón, J.%
, Carazo-Barbero, G.%
, Besada-Portas, E.%
\BCBL {}\ \BBA {} López-Orozco, J\BPBI A.%
\end{APACrefauthors}%
\unskip\
\newblock
\APACrefYearMonthDay{2023}{}{}.
\newblock
{\BBOQ}\APACrefatitle {Simulation-driven engineering for the management of
  harmful algal and cyanobacterial blooms} {Simulation-driven engineering for
  the management of harmful algal and cyanobacterial blooms}.{\BBCQ}
\newblock
\APACjournalVolNumPages{SIMULATION}{}{}{}.
\newblock
\begin{APACrefDOI} \doi{10.1177/00375497231184246} \end{APACrefDOI}
\PrintBackRefs{\CurrentBib}

\bibitem [\protect \citeauthoryear {%
Risco-Martín%
, Henares%
, Mittal%
, Almendras%
\BCBL {}\ \BBA {} Olcoz%
}{%
Risco-Martín%
, Henares%
\BCBL {}\ \protect \BOthers {.}}{%
{\protect \APACyear {2022}}%
}]{%
J-RiscoMartin2022}
\APACinsertmetastar {%
J-RiscoMartin2022}%
\begin{APACrefauthors}%
Risco-Martín, J\BPBI L.%
, Henares, K.%
, Mittal, S.%
, Almendras, L\BPBI F.%
\BCBL {}\ \BBA {} Olcoz, K.%
\end{APACrefauthors}%
\unskip\
\newblock
\APACrefYearMonthDay{2022}{}{}.
\newblock
{\BBOQ}\APACrefatitle {A Unified Cloud-Enabled Discrete Event Parallel and
  Distributed Simulation Architecture} {A unified cloud-enabled discrete event
  parallel and distributed simulation architecture}.{\BBCQ}
\newblock
\APACjournalVolNumPages{Simulation Modelling Practice and Theory}{}{}{}.
\newblock
\begin{APACrefDOI} \doi{10.1016/j.simpat.2022.102539} \end{APACrefDOI}
\PrintBackRefs{\CurrentBib}

\bibitem [\protect \citeauthoryear {%
Risco-Martín%
, Mittal%
, Henares%
, Cárdenas%
\BCBL {}\ \BBA {} Arroba%
}{%
Risco-Martín%
, Mittal%
\BCBL {}\ \protect \BOthers {.}}{%
{\protect \APACyear {2022}}%
}]{%
J-RiscoMartin2022b}
\APACinsertmetastar {%
J-RiscoMartin2022b}%
\begin{APACrefauthors}%
Risco-Martín, J\BPBI L.%
, Mittal, S.%
, Henares, K.%
, Cárdenas, R.%
\BCBL {}\ \BBA {} Arroba, P.%
\end{APACrefauthors}%
\unskip\
\newblock
\APACrefYearMonthDay{2022}{}{}.
\newblock
{\BBOQ}\APACrefatitle {x{DEVS}: {A} toolkit for interoperable modeling and
  simulation of formal discrete event systems} {x{DEVS}: {A} toolkit for
  interoperable modeling and simulation of formal discrete event
  systems}.{\BBCQ}
\newblock
\APACjournalVolNumPages{Softw: Pract and Exper}{}{}{}.
\newblock
\begin{APACrefDOI} \doi{10.1002/spe.3168} \end{APACrefDOI}
\PrintBackRefs{\CurrentBib}

\bibitem [\protect \citeauthoryear {%
Sengupta%
\ \BBA {} Andreas%
}{%
Sengupta%
\ \BBA {} Andreas%
}{%
{\protect \APACyear {2010}}%
}]{%
oahu}
\APACinsertmetastar {%
oahu}%
\begin{APACrefauthors}%
Sengupta, M.%
\BCBT {}\ \BBA {} Andreas, A.%
\end{APACrefauthors}%
\unskip\
\newblock
\APACrefYearMonthDay{2010}{March}{}.
\newblock
\APACrefbtitle {{Oahu Solar Measurement Grid (1-Year Archive): 1-Second Solar
  Irradiance; Oahu, Hawaii (Data)}.} {{Oahu Solar Measurement Grid (1-Year
  Archive): 1-Second Solar Irradiance; Oahu, Hawaii (Data)}.}
\newblock
\APAChowpublished {https://midcdmz.nrel.gov/apps/sitehome.pl?site=OAHUGRID}.
\newblock
\APACrefnote{Type: dataset (last accessed 22/6/2023)}
\newblock
\begin{APACrefDOI} \doi{10.5439/1052451} \end{APACrefDOI}
\PrintBackRefs{\CurrentBib}

\bibitem [\protect \citeauthoryear {%
Sisodia%
\ \BBA {} Mathur%
}{%
Sisodia%
\ \BBA {} Mathur%
}{%
{\protect \APACyear {2019}}%
}]{%
Sisodia2019}
\APACinsertmetastar {%
Sisodia2019}%
\begin{APACrefauthors}%
Sisodia, A\BPBI K.%
\BCBT {}\ \BBA {} Mathur, R\BPBI K.%
\end{APACrefauthors}%
\unskip\
\newblock
\APACrefYearMonthDay{2019}{}{}.
\newblock
{\BBOQ}\APACrefatitle {Impact of bird dropping deposition on solar photovoltaic
  module performance: a systematic study in Western Rajasthan} {Impact of bird
  dropping deposition on solar photovoltaic module performance: a systematic
  study in western rajasthan}.{\BBCQ}
\newblock
\APACjournalVolNumPages{Environmental Science and Pollution
  Research}{26}{30}{31119--31132}.
\PrintBackRefs{\CurrentBib}

\bibitem [\protect \citeauthoryear {%
T.%
, Bharamagoudar%
, G.%
\BCBL {}\ \BBA {} Totad%
}{%
T.%
\ \protect \BOthers {.}}{%
{\protect \APACyear {2021}}%
}]{%
Prophet}
\APACinsertmetastar {%
Prophet}%
\begin{APACrefauthors}%
T., N.%
, Bharamagoudar, G\BPBI R.%
, G., K\BPBI K.%
\BCBL {}\ \BBA {} Totad, S\BPBI G.%
\end{APACrefauthors}%
\unskip\
\newblock
\APACrefYearMonthDay{2021}{}{}.
\newblock
{\BBOQ}\APACrefatitle {Real-Time Anomaly Detection Using Facebook Prophet}
  {Real-time anomaly detection using facebook prophet}.{\BBCQ}
\newblock
\APACjournalVolNumPages{International Journal of Natural Computing
  Research}{10}{3}{29–40}.
\newblock
\begin{APACrefDOI} \doi{10.4018/IJNCR.2021070103} \end{APACrefDOI}
\PrintBackRefs{\CurrentBib}

\bibitem [\protect \citeauthoryear {%
Toharudin%
\ \protect \BOthers {.}}{%
Toharudin%
\ \protect \BOthers {.}}{%
{\protect \APACyear {2023}}%
}]{%
Toharudin2023}
\APACinsertmetastar {%
Toharudin2023}%
\begin{APACrefauthors}%
Toharudin, T.%
, Pontoh, R\BPBI S.%
, Caraka, R\BPBI E.%
, Zahroh, S.%
, Lee, Y.%
\BCBL {}\ \BBA {} Chen, R\BPBI C.%
\end{APACrefauthors}%
\unskip\
\newblock
\APACrefYearMonthDay{2023}{}{}.
\newblock
{\BBOQ}\APACrefatitle {Employing long short-term memory and Facebook prophet
  model in air temperature forecasting} {Employing long short-term memory and
  facebook prophet model in air temperature forecasting}.{\BBCQ}
\newblock
\APACjournalVolNumPages{Communications in Statistics-Simulation and
  Computation}{52}{2}{279--290}.
\PrintBackRefs{\CurrentBib}

\bibitem [\protect \citeauthoryear {%
Vishwas%
\ \BBA {} Patel%
}{%
Vishwas%
\ \BBA {} Patel%
}{%
{\protect \APACyear {2020}}%
}]{%
Vishwas2020}
\APACinsertmetastar {%
Vishwas2020}%
\begin{APACrefauthors}%
Vishwas, B\BPBI V.%
\BCBT {}\ \BBA {} Patel, A.%
\end{APACrefauthors}%
\unskip\
\newblock
\APACrefYear{2020}.
\newblock
\APACrefbtitle {Hands-on Time Series Analysis with Python: From Basics to
  Bleeding Edge Techniques} {Hands-on time series analysis with python: From
  basics to bleeding edge techniques}.
\newblock
\APACaddressPublisher{}{Apress}.
\PrintBackRefs{\CurrentBib}

\bibitem [\protect \citeauthoryear {%
Wen%
\ \protect \BOthers {.}}{%
Wen%
\ \protect \BOthers {.}}{%
{\protect \APACyear {2019}}%
}]{%
STL}
\APACinsertmetastar {%
STL}%
\begin{APACrefauthors}%
Wen, Q.%
, Gao, J.%
, Song, X.%
, Sun, L.%
, Xu, H.%
\BCBL {}\ \BBA {} Zhu, S.%
\end{APACrefauthors}%
\unskip\
\newblock
\APACrefYearMonthDay{2019}{}{}.
\newblock
{\BBOQ}\APACrefatitle {RobustSTL: A Robust Seasonal-Trend Decomposition
  Algorithm for Long Time Series} {Robuststl: A robust seasonal-trend
  decomposition algorithm for long time series}.{\BBCQ}
\newblock
\APACjournalVolNumPages{Proceedings of the AAAI Conference on Artificial
  Intelligence}{33}{01}{5409-5416}.
\PrintBackRefs{\CurrentBib}

\bibitem [\protect \citeauthoryear {%
D.~Yang%
}{%
D.~Yang%
}{%
{\protect \APACyear {2019}}%
}]{%
Yang2019}
\APACinsertmetastar {%
Yang2019}%
\begin{APACrefauthors}%
Yang, D.%
\end{APACrefauthors}%
\unskip\
\newblock
\APACrefYearMonthDay{2019}{}{}.
\newblock
{\BBOQ}\APACrefatitle {A guideline to solar forecasting research practice:
  Reproducible, operational, probabilistic or physically-based, ensemble, and
  skill (ROPES)} {A guideline to solar forecasting research practice:
  Reproducible, operational, probabilistic or physically-based, ensemble, and
  skill (ropes)}.{\BBCQ}
\newblock
\APACjournalVolNumPages{Journal of Renewable and Sustainable
  Energy}{11}{2}{022701}.
\PrintBackRefs{\CurrentBib}

\bibitem [\protect \citeauthoryear {%
H\BHBI T.~Yang%
, Huang%
, Huang%
\BCBL {}\ \BBA {} Pai%
}{%
H\BHBI T.~Yang%
\ \protect \BOthers {.}}{%
{\protect \APACyear {2014}}%
}]{%
Yang2014}
\APACinsertmetastar {%
Yang2014}%
\begin{APACrefauthors}%
Yang, H\BHBI T.%
, Huang, C\BHBI M.%
, Huang, Y\BHBI C.%
\BCBL {}\ \BBA {} Pai, Y\BHBI S.%
\end{APACrefauthors}%
\unskip\
\newblock
\APACrefYearMonthDay{2014}{}{}.
\newblock
{\BBOQ}\APACrefatitle {A weather-based hybrid method for 1-day ahead hourly
  forecasting of PV power output} {A weather-based hybrid method for 1-day
  ahead hourly forecasting of pv power output}.{\BBCQ}
\newblock
\APACjournalVolNumPages{IEEE transactions on sustainable
  energy}{5}{3}{917--926}.
\PrintBackRefs{\CurrentBib}

\bibitem [\protect \citeauthoryear {%
Zeigler%
, Muzy%
\BCBL {}\ \BBA {} Kofman%
}{%
Zeigler%
\ \protect \BOthers {.}}{%
{\protect \APACyear {2018}}%
}]{%
Zeigler2018}
\APACinsertmetastar {%
Zeigler2018}%
\begin{APACrefauthors}%
Zeigler, B\BPBI P.%
, Muzy, A.%
\BCBL {}\ \BBA {} Kofman, E.%
\end{APACrefauthors}%
\unskip\
\newblock
\APACrefYear{2018}.
\newblock
\APACrefbtitle {Theory of modeling and simulation: discrete event \& iterative
  system computational foundations} {Theory of modeling and simulation:
  discrete event \& iterative system computational foundations}.
\newblock
\APACaddressPublisher{}{Academic press}.
\PrintBackRefs{\CurrentBib}

\bibitem [\protect \citeauthoryear {%
Zhang%
, Kleiber%
, Florita%
, Hodge%
\BCBL {}\ \BBA {} Mather%
}{%
Zhang%
\ \protect \BOthers {.}}{%
{\protect \APACyear {2018}}%
}]{%
Zhang2018}
\APACinsertmetastar {%
Zhang2018}%
\begin{APACrefauthors}%
Zhang, W.%
, Kleiber, W.%
, Florita, A\BPBI R.%
, Hodge, B\BHBI M.%
\BCBL {}\ \BBA {} Mather, B.%
\end{APACrefauthors}%
\unskip\
\newblock
\APACrefYearMonthDay{2018}{}{}.
\newblock
{\BBOQ}\APACrefatitle {Modeling and simulation of high-frequency solar
  irradiance} {Modeling and simulation of high-frequency solar
  irradiance}.{\BBCQ}
\newblock
\APACjournalVolNumPages{IEEE Journal of Photovoltaics}{9}{1}{124--131}.
\PrintBackRefs{\CurrentBib}

\end{thebibliography}
